\documentclass[12pt]{article}
\usepackage{epsfig}
\usepackage{amsmath}

\makeatletter
\@addtoreset{equation}{section}

\newcommand{\p}{\partial}

\newcommand{\mh}{\hat{\mu}}
\newcommand{\nh}{\hat{\nu}}
\def \be {\begin{equation}}
\def \ee {\end{equation}}
\def \bea {\begin{eqnarray}}
\def \eea {\end{eqnarray}}

\def \del {\partial}
\def \dels {\partial\kern-.5em / \kern.5em}
\def \As {{A\kern-.5em / \kern.5em}}
\def \Ds {D\kern-.7em / \kern.5em}

\def \Lag {\tilde{\cal L}}

\def \Dbar {{$\bar{\mbox{D}}$}}

\begin{document}


\begin{titlepage}
\begin{center}

\hfill\parbox{4cm}{
{\normalsize\tt hep-th/0303172}\\
{\normalsize UT-Komaba/03-4}
}

\vskip 1in

{\LARGE \bf Time Evolution via S-branes}

\vskip 0.3in

{\large
Koji {\sc Hashimoto},\hspace{-3pt}$^a$\footnote{\tt
  koji@hep1.c.u-tokyo.ac.jp} \
Pei-Ming {\sc Ho},\hspace{-3pt}$^b$\footnote{{\tt
  pmho@phys.ntu.edu.tw}} \
Satoshi {\sc Nagaoka}$^a$\footnote{{\tt
 nagaoka@hep1.c.u-tokyo.ac.jp}}
and \\
John E.\ {\sc Wang}$^{b}$\footnote{{\tt
hllywd2@phys.ntu.edu.tw}} }

\vskip 0.15in

${}^a$ {\it Institute of Physics, University of Tokyo, Komaba}\\
{\it Tokyo 153-8902, Japan}\\[3pt]
${}^b$ {\it Department of Physics, National Taiwan University \\
Taipei 106, Taiwan}\\
[0.3in]

{\normalsize March 2003}

\end{center}

\vskip .3in

\begin{abstract}
\normalsize\noindent
Using S(pacelike)-branes defined through rolling tachyon
solutions, we show how the dynamical formation of
D(irichlet)-branes and strings in tachyon condensation can be
understood. Specifically we present solutions of S-brane actions
illustrating the classical confinement of electric and magnetic
flux into fundamental strings and D-branes. The role of S-branes
in string theory is further clarified and their RR charges are
discussed. In addition, by examining ``boosted'' S-branes, we
find what appears to be a surprising dual S-brane description of
strings and D-branes, which also indicates that the critical
electric field can be considered as a self-dual point in string
theory. We also introduce new tachyonic S-branes
as Euclidean counterparts to non-BPS branes.

\end{abstract}

\vfill

\end{titlepage}
\setcounter{footnote}{0}

\pagebreak
\renewcommand{\thepage}{\arabic{page}}
{\baselineskip=5mm\tableofcontents}

\section{Introduction}

Tachyon condensation in open string theories has revealed new
intriguing aspects of string theories and D-branes. One of the
meritorious achievements in this area is that we can now describe
D-branes as topological solitons in (effective) field theories of
tachyons and string field theories.  This approach to D-branes has
also been extended to deal with the time dependent decay or
creation of D-branes.  In developing tools to deal with the
complexities of time dependent systems, new string theory
ingredients called S(pacelike)-branes were introduced in
Ref.~\cite{stro}. Whereas ordinary D-branes are realized as
timelike kinks and vortices of the tachyon field, spacelike
defects can be defined as spacelike kinks and vortices in the
background of a time dependent tachyon condensation process
called rolling tachyons \cite{roll}. As defined S-branes are
intrinsically related to and naturally arise in time dependent
processes in string theory.\footnote{See
Refs.~\cite{followS,J7,Sentimeevole,Strotalk,LNT,IshidaUehara,MSY}
for the development following Ref.~\cite{stro}.  Early work on
tachyon condensation includes Ref.~\cite{Halpern}.}

In Ref.~\cite{Sbraneaction}, part of the present authors
demonstrated that S-branes can in fact describe the formation of
topological defects in time dependent tachyon condensation.  The
key point was that while flat S-branes are defined as spacelike
defects of a specific rolling tachyon solution, we can also
introduce fluctuations into the rolling tachyon which will
accordingly deform the S-branes.  It was then found that the
information from only the S-brane fluctuations is sufficient to
describe the formation of individual fundamental strings as
remnants of the original tachyon system. The advantage of the
S-brane approach in describing tachyon remnant formation came
from the fact that explicit knowledge of the full tachyon action
was not necessary. This is a generalized correspondence between
tachyon systems and Dirac-Born-Infeld (DBI) systems on the
tachyon defects \cite{tsey, hirano}. S-branes are universally
governed by a Euclidean DBI effective action, independent of the
specific details of the original tachyon systems, and with scalar
excitations along the time direction.  While many tachyonic
Lagrangians have similar features and give rise to the same type
of static solitons and rolling tachyon backgrounds, we must look
for these solutions in each Lagrangian individually.  Another
advantage of the S-brane approach is then that an S-brane
solution represents a class of solutions for many tachyonic
Lagrangians; these solutions are classes in the sense that many
different tachyonic Lagrangians give rise to the same type of
S-brane solutions. So while in string theory the tachyon
effective actions are obtained in various forms with different
derivations, the S-brane approach gives a universal treatment.
A third advantage is that it is easier to solve the equations of
motion for the S-brane action than for arbitrary tachyon systems.

In this paper, after discussing S-branes and their role in time
dependent physics in Section \ref{Sbranerole}, we will illustrate
our ideas by presenting classical solutions of the S-brane
actions, clarifying their role and obtaining their corresponding
tachyon descriptions.\footnote{We neglect closed string
backreactions when describing the rolling tachyon.} In Section
\ref{sfs} we recapitulate the solution \cite{Sbraneaction} of the
formation of confined electric fluxes which are fundamental
strings. In addition we show how the S-brane solution is
consistent with the tachyon picture of classical flux
confinement.  In Section~\ref{dfs} new solutions representing the
formation of $(p,q)$ strings are presented and we relate these new
solutions to an implementation of S-duality for S-branes. The
late time behavior of these S-brane solutions can be captured by
simple linear solutions which we call ``boosted'' S-branes. These
boosted S-branes are given corresponding explicit tachyon
solutions and boundary state descriptions in Section~\ref{bsas},
and their consistency with the usual string and D-brane picture
is checked. T-duality in the time direction is found to
interchange these two classes of D-brane solutions with the
electric field above or below the critical value. In Section
\ref{sadi} we examine the possibility that S-brane solutions may
describe D-brane scattering and Feynman diagrams for D-branes. We
further find a generalized RR charge conservation law for
S-/D-branes. Section \ref{cad} is devoted to conclusions and
discussions.

It should be emphasized that although we are using the language
of string theory, any theory with topological defects will have
its own ``S-branes'' or spacelike defects. Some of these
solutions should necessarily describe defect formation. It would
be fascinating if our methods can be further applied to the
formation of other topological defects and also provide dual
descriptions of all kinds of defects and remnants.

In the paper we take $2\pi\alpha'=1$ unless stated otherwise.

\section{Roles of S-branes}\label{Sbranerole}

The central idea we explore throughout this paper is how S-branes
can be used to describe time dependent defect formation and
tachyon condensation decay remnants. The detailed exploration of
the classical solutions of S-brane actions will be provided in
later sections, and we first concentrate on general properties of
S-branes, explaining their important roles in time-dependent
tachyon condensation. Along the way we will see how S-branes and
their classical solutions can be classified by the species of
tachyon remnants, and discuss a new type of S-brane, which we name
tachyonic S-brane.  We also derive S-brane actions which have a
universal form, slightly generalizing the results in
Ref.~\cite{Sbraneaction}.

\subsection{Remnant or defect formation}

Assuming that the tachyon potential for
a non-BPS D-brane is minimized
at some values for both $T>0$ and $T<0$,
kink solutions can be approximately depicted by
the $T = 0$ loci.
While the timelike kinks correspond to D-branes,
the spacelike ones are S-branes.
When S-branes were first introduced,
they provided a fresh approach
to the study of time dependent systems,
but only fine tuned configurations were considered.
Actually, as we will now demonstrate,
S-branes appear ubiquitously during tachyon condensation.
This is why it is worthwhile to define
the S-brane action and to study its general solutions
\cite{Sbraneaction}.

\begin{figure}[bhtp]
\begin{center}
\includegraphics[width=13cm]{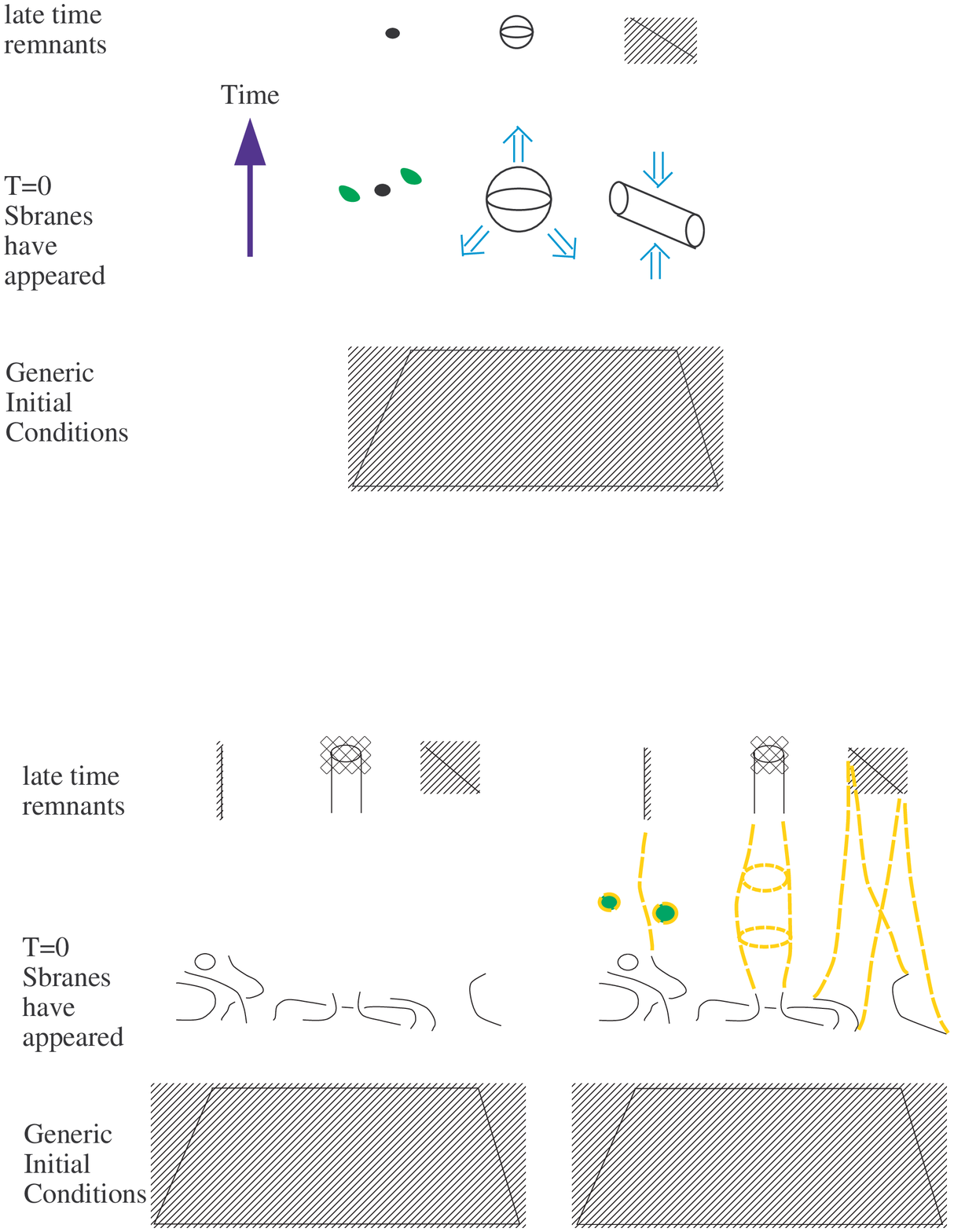}
\caption{The top figure is a series of snapshots of tachyon time
evolution processes but since time is not explicit, the role of
the S-brane is obscured.  The bottom left figure is essentially
just a redrawing of the top figure.  The bottom right figure
shows the entire dynamical evolution process with the S-branes
outlined.  The $T=0$ regions are drawn in as dashed lines. The
main point is that at late times we have remnants with tachyon
value zero and we can produce them from generic initial
conditions. S-branes are how we ``connect the lines'' from the
initial to final stage.  } \label{Sbrane-spacetimecombo}
\end{center}
\end{figure}

At late times of the tachyon condensation process, it is possible
to describe D-brane remnants as kinks (or lumps) in the tachyon
potential.
In principle it should be possible to follow the
time evolution of these $T=0$ regions.
One might ask why
we need to consider S-branes.
The point is that,
given a generic tachyon configuration,
before the remnants are fully formed
(before the tachyon profiles are localized),
S-branes appear first
in the time dependent formation of defects.
These $T=0$ regions
can ``appear out of nowhere'' at some time and are exactly
S-branes.
Only when the $T\simeq 0$ region becomes spatially localized,
has the S-brane metamorphosed or
decayed into a D-brane (topological defect), see
Fig.~\ref{Sbrane-spacetimecombo}.
In addition, even if there are no remnants,
short-lived S-branes will appear as long
as the energy is large enough to create local fluctuations over
the top of the tachyon potential.

Furthermore, although it is suggested by its name
and usually assumed that the S-branes are spacelike,
the S-brane action admits timelike solutions
which correspond to D-branes with a large electric field.
We have seen such solutions in Ref.~\cite{Sbraneaction}
and will present others below.

\subsection{S-branes as classes of tachyon decay}

In the case of tachyonic Lagrangians, it is possible to find kink
solutions which represent lower dimensional excitations such as
D-branes. These relations between unstable branes and ``static''
branes are also called the descent relations.  A different
question one can ask is how are the various objects in string
theory related when we take into account time dependent
processes? If we start off with a tachyonic system and end up
with a stable system, then what is the time evolution process
which connects these two systems?  We propose that S-branes be
used to classify the time evolution processes whenever there are
remnants in the end.

\begin{figure}[htp]
\begin{center}
\includegraphics[width=10cm]{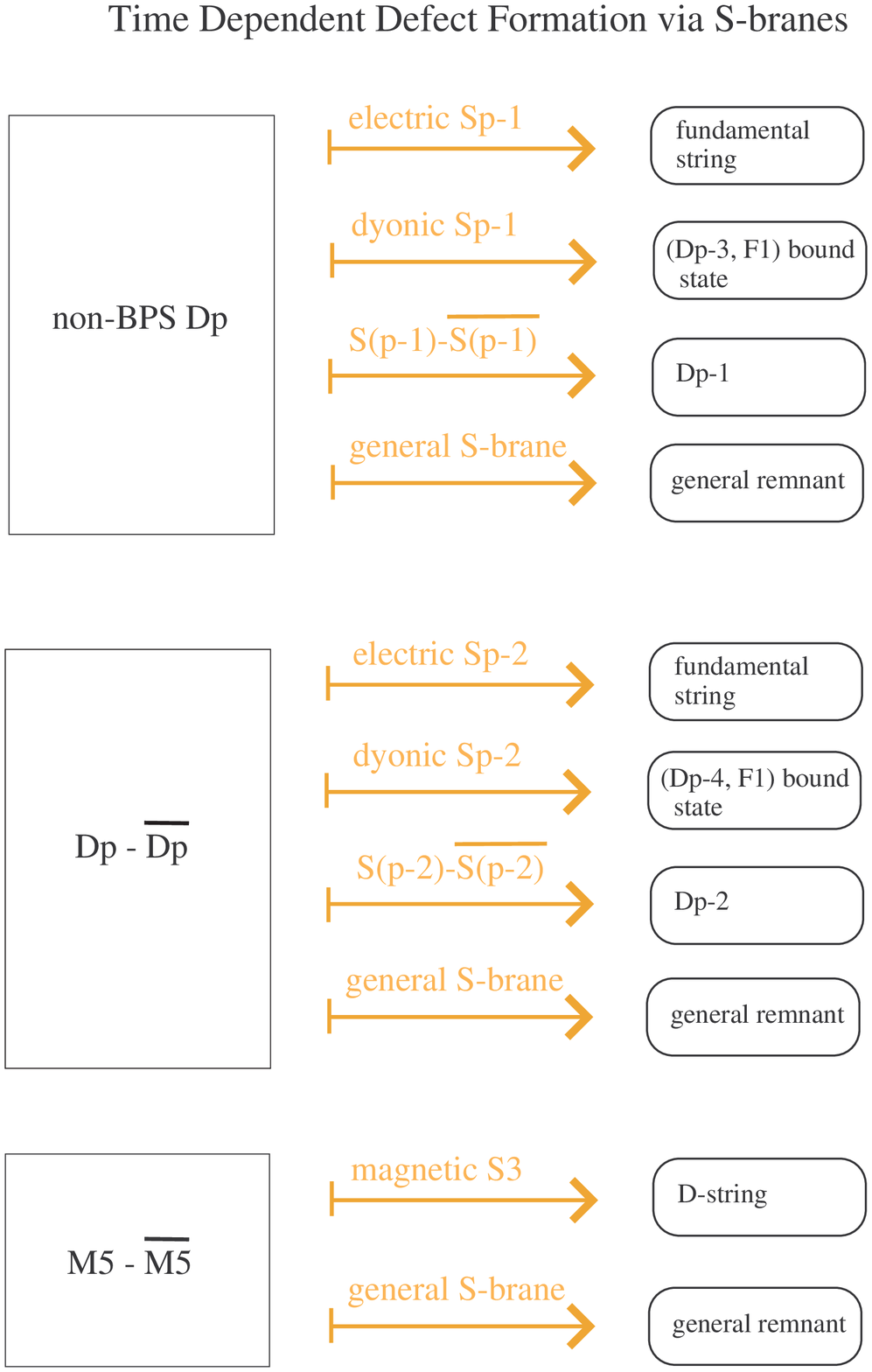} \caption{Time
evolution processes characterized by S-branes. The S-branes are
the arrows. The upper three arrows starting from the non-BPS
D$p$-brane will be treated in Sections 3,4 and 6 respectively.
Although the S-branes from the non-BPS brane basically have
counterparts in the D$p$-$\overline{{\rm D}p}$, the arrows
emanating from the D$p$-$\overline{{\rm D}p}$ include processes
previously unknown, especially the ones mediated by tachyonic
S-branes.  All arrows are commonly expected both in type IIA and
IIB string theories.  Finally, to understand the creation of
D-strings, it is necessary to incorporate M-theory effects as
indicated in the bottom figure and discussed in Section 4.
}
\label{remnantchart}
\end{center}
\end{figure}

We emphasize that there are differences between the S-branes of
the non-BPS brane and the D-\Dbar\ system. It is clear that the
S-branes share common properties but there should also be some
differences due to the additional tachyon on the D-\Dbar\ pair.
There are additional S-branes for the D-\Dbar\ system which we
call ``tachyonic S-branes'' which might be considered Euclidean
counterparts of non-BPS branes, in view of the correspondence
that the original S-branes are Euclidean counterparts of BPS
D-branes; the precise correspondence between tachyonic S-branes
and Euclidean non-BPS branes is however not clear (see the next
subsection for the precise definition of the tachyonic S-branes).
Tachyonic S-branes should not be hard
to differentiate from S-branes and describe essentially different
time evolution processes. Some processes might be solutions of
S-brane Lagrangians and some might be solutions of tachyonic
S-brane Lagrangians.  With this point in mind, we summarize the
solutions discussed in this paper in Fig.~\ref{remnantchart}.

\begin{figure}[htp]
\begin{center}
\begin{minipage}{14cm}
\begin{center}
\includegraphics[width=14cm]
{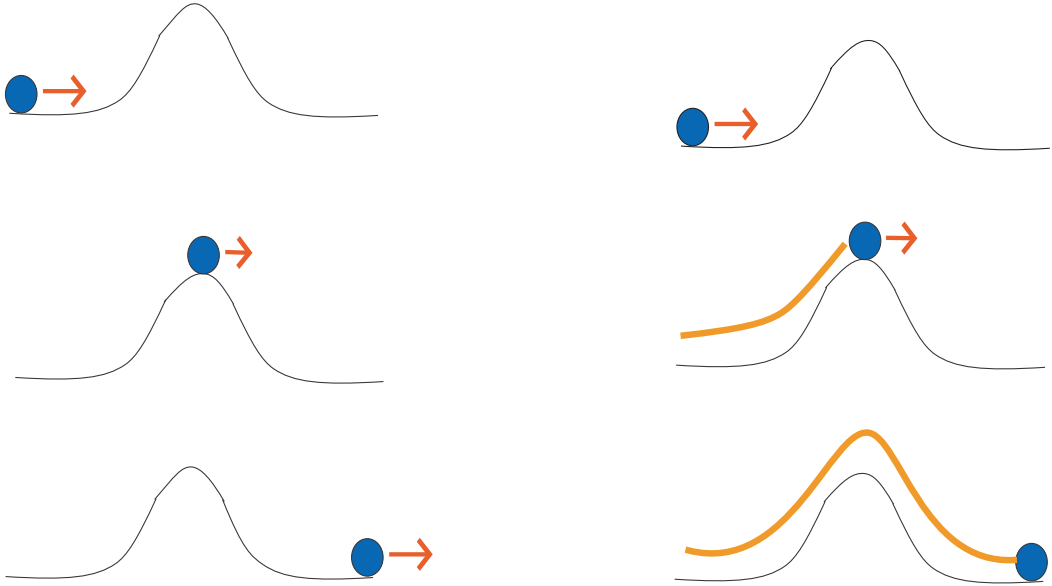} \caption{Two different time evolution
processes characterized by S-branes.  The three pictures on the
left characterize the rolling tachyon picture so the S-brane
appears only when the tachyon crosses the top of the potential.
The second three pictures give a schematic of remnant creation. We
start off with some energy in the tachyon and perhaps in other
fields.  As the tachyon rolls, at some point it starts to create
$T=0$ regions specified by the thick lines which eventually turn
into remnants. At late times, the tachyon does not roll (no
velocity arrow) as all the energy has been transferred into the
remnant kink.} \label{pullingremnant}
\end{center}
\end{minipage}
\end{center}
\end{figure}

In Ref.~\cite{stro} S-branes represented a tachyon configuration
rolling up and down the tachyon potential with the energy
necessary to go up the potential remaining as some background
contribution. This means at late times we have a time evolving
system with energy stored in either radiation, the rolling
tachyon or various other fields. In our case, however, long lived
S-branes represent remnant formation and this difference implies
that the process is not always time reversal invariant.
As an example of the process we are considering, let us consider
a finite energy configuration with the tachyon at large negative
values.  As the system evolves we climb up the tachyon potential,
and at some point an S-brane shows up and eventually creates a
remnant.  The energy of the configuration can then be totally
transferred to the remnant, so the S-brane shows how delocalized
systems organize and transform energy into a remnant; in the end
there might be no energy left to go into radiation, rolling
tachyon or anything else.\footnote{ In the argument here we
compactify directions transverse to the resultant remnant in the
worldvolume of the original unstable brane. This is necessary for
the remnant to possess a finite tension. This observation is
consistent with what has been studied in other literature
\cite{Sentimeevole,LNT,IshidaUehara}. } The S-brane schematically
pulls the tachyon values over the potential and leaves a remnant
solution in the process, see Figs.~\ref{pullingremnant} and
\ref{tachyonvalues}.

Ref.~\cite{stro} also discusses the width of an S-brane. In the
context of tachyon condensation an analogous question is how easy
is it to put one flat S-branes one after another in time. In
general it is not clear if there is some limiting factor since it
takes time for the tachyon to roll up and down the potential,
however it should not be impossible to have multiple S-branes.
Any initial conditions forming the rolling tachyon can simply be
repeated at some later time so this will roughly produce two
separated rolling tachyon processes and two flat S-branes.  It is
the interactions between the initial conditions which will place a
limit on how easy it is to produce multiple S-branes.  This
question could be explored further and it is related to
coincident S-branes and their possible non-Abelian structure.

\subsection{S-brane descent relations and new ``tachyonic'' S-branes}
\label{vortex}

\begin{figure}[tp]
\begin{center}
\begin{minipage}{13cm}
\begin{center}
\includegraphics[width=10cm]{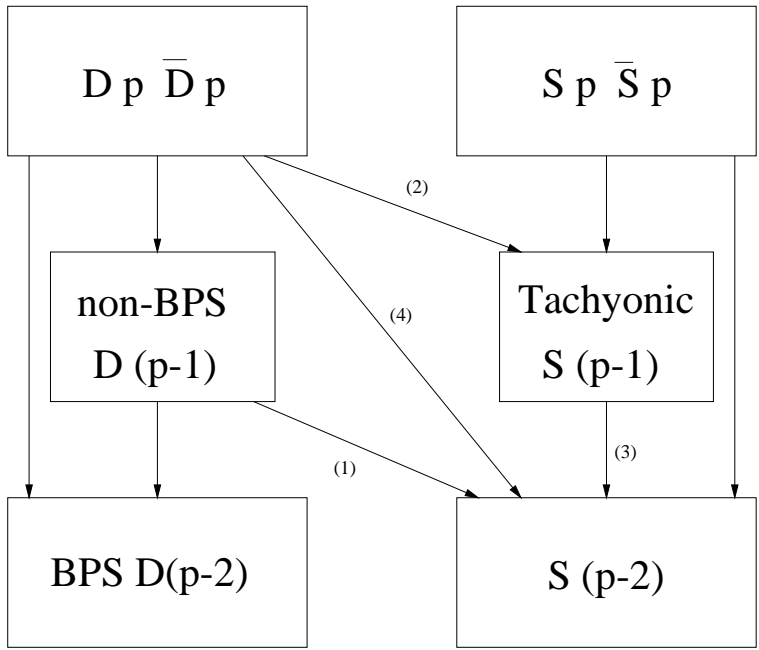}
\end{center}
\caption{The extended descent relation for tachyon condensations.
We do not deal with the relation between type IIA and type IIB
here. }
\label{descentt}
\end{minipage}
\end{center}
\end{figure}

It has been argued that static tachyonic kink solutions on non-BPS
branes correspond to codimension one BPS branes, while vortex
solutions on D-\Dbar\ pairs are codimension two BPS branes. The
relationship between these branes is summarized by the usual
descent relations \cite{SenTachyon}. In analogy, Gutperle and
Strominger \cite{stro} also defined S-branes as time-dependent
kinks (vortices) on non-BPS branes (D-\Dbar\ pairs), so it should
be possible to extend the descent relations, shown in
Fig.~\ref{descentt}, to include both D-branes and S-branes. One
may understand that the horizontal correspondence in the figure
is just Euclideanization, or the change ``timelike
$\leftrightarrow $ spacelike''.  For example, from this view
point the relation between the S$(p\!-\!2)$-brane and the non-BPS
D$(p\!-\!1)$-brane can be understood\footnote{ Note that the
arrows in this figure are not the physical processes of formation
which are depicted in Fig.~\ref{remnantchart}. Here the arrows
just represent construction of classical solutions from
Lagrangians. } as an arrow (1) in the extended descent relations.
This arrow is how one can derive an S-brane action from the
non-BPS D-brane action \cite{Sbraneaction}.  The $D(p-2)$ vortex
solution on a D$p$-$\bar{\rm D}p$ can be generalized to an S-brane
counterpart.  Later in this section we will derive the action of
an S-brane spacetime vortex along the arrow (4).

First, starting at the top right of Fig.~\ref{descentt} we have
an S-$\bar{\rm S}$ pair. The figure also contains the tachyonic
S$(p\!-\!1)$-brane.  The tachyonic brane is naturally embedded
into the extended descent relation since the space-time vortex
(the arrow (4) in the figure) from D-$\bar{\rm D}$ to an S$(p-2)$
can be decomposed into two procedures : first construct a
time-dependent kink (2) and then a space-dependent kink (3). The
second procedure is almost the same as the arrow from the non-BPS
D$(p-1)$ to the BPS D$(p-2)$.

To understand what a tachyonic S-brane is, let us first construct it.
We begin with the Lagrangian of a D$p$-$\overline{{\rm D}p}$
pair, choosing
the Lagrangian of the boundary string field theory (BSFT)
\cite{BSFT,BSFTsoliton,BSFTRR}
since it is the best understood. The recent
paper by Jones and Tye \cite{JT} proposed the action
\begin{eqnarray}
 S = -{2T_{\rm  D9}} \int d^{10} x \
e^{-\pi |T|^2} {\cal F}(X + \sqrt{Y}){\cal F}(X - \sqrt{Y}),
\label{BSFT}
\end{eqnarray}
where we define $X \equiv \p_\mu T \p^\mu \bar{T}$ and $Y \equiv
(\p_\mu T)^2(\p^\nu \bar{T})^2$, and for simplicity we choose
$p=9$. We don't need detailed information of the kinetic function
${\cal F}$ here. This action is valid for linear tachyon
profiles, but unfortunately a linear ansatz for time-dependent
homogeneous solutions $T=T(x^0)$ leads to only trivial solutions
(see Ref.~\cite{BSFTmatter}). Even though we exceed the validity
of the action, let us proceed for the moment and examine the
homogeneous tachyon solution. Noting that the D-\Dbar\ system
reduces to the non-BPS brane system when we restrict the complex
tachyon $T=T_1 + iT_2$ to take only real value $T_1$, it is easy
to see that the classical solution presented in
Ref.~\cite{BSFTmatter},
\begin{eqnarray}
 T = T_{\rm cl}(x^0) = x^0 +
\mbox{[exponentially small terms for large $x^0$]},
\label{solts}
\end{eqnarray}
is the tachyon solution on the D-\Dbar\ which we are looking for.
The imaginary part $T_2$ of the complex tachyon appears in the
Lagrangian only in squared form and so the equation of motion for
$T_2$ has an overall factor $T_2$ or $\p T_2$ and is trivially
satisfied by $T_2=0$.  However the ``tachyonic'' fluctuation from
$T_2$ leads to a new feature which we call
the tachyonic S-brane. An effective
tachyonic S-brane action is discussed in Appendix \ref{appa}.

Next, we consider the arrow (4) in this subsection, which
will provide another way to derive the S-brane action.
This solution can be thought of as a
combination of a time dependent kink
and the usual space-dependent kink along $x^1$.
The solution of the
BSFT action (\ref{BSFT}) is easily found
\begin{eqnarray}
 T = T_{\rm cl}(x^0) + i u x^1
\end{eqnarray}
where $u$ goes to infinity by the usual BSFT argument for
spatial kinks \cite{BSFTsoliton,BSFTRR}.
This classical solution has two zero modes in
fluctuations since
this ``space-time vortex''
breaks two translation symmetries.

Following the analysis of Ref.~\cite{KojiSatoshi-Descent} we
construct an effective action of the space-time vortex which we
identify as an S-brane.  The effective action of a D9-$\bar{\rm
{D}}$9 system takes the form
\begin{eqnarray}
S=2 T_{\rm{D9}} \int d^{10} x \ e^{-\pi |T|^2} \sqrt{\det
(1+F)}f(X,Y)
\end{eqnarray}
where $F$ is the diagonal linear combination of the two $U(1)$
gauge fields, $F=F_1+F_2$ and $X,Y$ are now defined using
the open string metric with respect for $F$
\begin{eqnarray}
X \equiv G^{\mu\nu}\p_\mu T \p_\nu \bar{T}, \quad Y \equiv
|G^{\mu\nu}\p_\mu T\p_\nu T|^2 .
\end{eqnarray}
This effective action is constrained by the usual assumption that
the fields are slowly varying.
The fluctuation fields which are zero modes (Nambu-Goldstone modes) are
embedded in the action in a special manner since it is associated
with the breaking of the translational symmetries. In fact, they
appear as a kind of Lorentz transformation,
\begin{eqnarray}
  T = T_{\rm  sol}(y_0, y_1),\quad
&&y_0 \equiv \frac{1}{\beta_0} \Bigl( x_0 - t_0(x_{\mh} )\Bigr),
\quad
y_1 \equiv \frac{1}{\beta_1} \Bigl( x_1 - t_1(x_{\mh})\Bigr)\\
&&x \to y = \Lambda x , \quad (\Lambda^t) G \Lambda = G,
\label{lorentz}
\end{eqnarray}
where the open string metric is (we turn on only
$F_{\hat{\mu}\hat{\nu}}$ ($\hat{\mu}, \hat{\nu}=2,\cdots,9$))
\begin{eqnarray}
G^{\mh\nh} =\left( \frac{1}{1-F^2} \right)^{\mh\nh},\quad G^{00}
= -1, \quad G^{11} = 1, \quad G^{0 \mh}=G^{1 \nh}=0
\end{eqnarray}
and the Lorentz transformation matrix $\Lambda$ is
\begin{eqnarray}
  \Lambda = \left(
    \begin{array}{cc|c}
1/\beta_0 & 0 & -\p_{\mh} t_0/\beta_0\\
0 & 1/\beta_1 & -\p_{\mh} t_1/\beta_1\\
\hline
{}* &* &* \\
\vdots &\vdots &\vdots
    \end{array}
\right).
\end{eqnarray}
Lorentz invariance (\ref{lorentz}) of the open string metric
determines the beta factors
\begin{eqnarray}
  \beta_0 = \sqrt{1- G^{\mh \nh}\p_{\mh} t_0 \p_{\nh} t_0},\quad
  \beta_1 = \sqrt{1+ G^{\mh\nh}\p_{\mh} t_1 \p_{\nh} t_1},\quad
 G^{\mh\nh}\p_{\mh} t_0 \p_{\nh} t_1=0
\end{eqnarray}
which can be substituted back into the action to give,
after performing the integration over $x^0$ and $x^1$,
\begin{eqnarray}
  S &=& S_0 \int d^8 x \beta_0 \beta_1 \sqrt{\det
    \left(\delta_{\mh\nh}+F_{\mh\nh}\right) } \nonumber \\
 &=& S_0 \int d^8 x \sqrt{\det
    \left(\delta_{\mh\nh}+ F_{\mh\nh}- \p_{\mh} t_0 \p_{\nh} t_0
+\p_{\mh} t_1 \p_{\nh} t_1  \right) }.  \label{vortexaction}
\end{eqnarray}
This is the effective action for the spacetime vortex, coinciding
with the S-brane action which was derived in
Ref.~\cite{Sbraneaction} if we set $t_1 = 0$. The new scalar
field $t_1$ appears in the same way as how the usual D-brane
action is generalized to the D-\Dbar\ pair. This action naturally
leads to the following general form of the S$p$-brane action in
which the worldvolume embedding in the bulk spacetime ($X^M$ with
$M=0,1,\cdots,9$) has not been gauge-fixed
\begin{eqnarray}
 S = S_0 \int \! d^{p+1}x \sqrt{{\rm det}
(\p_{\hat{\mu}} X_M \p_{\hat{\nu}} X^M + F_{\hat{\mu}\hat{\nu}}
)} \ .
\end{eqnarray}
The field $t_0$ in Eq.~(\ref{vortexaction}) is identified with
the embedding scalar $X^0$. Since in our derivation we did not
refer to a specific tachyon effective action, the form of the
S-brane action is universal in the slowly-varying field
approximation.\footnote{We expect that our S-brane action derived
using a field theoretic approach is related to the long-distance
S-brane effective field theory in Ref.~\cite{MSY}.}


\section{Strings from S-branes}
\label{sfs}

S-brane solutions describing a flux tube confining into a
fundamental string have been previously discussed in
Ref.~\cite{Sbraneaction}. In this section we re-examine the
solution from a spacetime perspective which will be helpful in
finding other S-brane solutions in the next section.  Also, by
directly analyzing the tachyon system, we find further evidence
that the S-brane solution should be regarded as a fundamental
string.

\subsection{Solution of F-string formation}

Let us review the electric S3-brane spike solution of
Ref.~\cite{Sbraneaction}.\footnote{Ref.~\cite{Sbraneaction}
discusses S$p$-branes with $p\geq 3$, but in this section we
consider the $p=3$ case in preparation of Section \ref{dfs}.} The
S-brane actions of Eq.~(\ref{vortexaction}) were derived in a
certain gauge in which the time direction was treated as a scalar
field $X^0$. In the following sections we will discuss S-brane
solutions with nontrivial time dependence, so we take the
following gauge choice which is preferable in the spacetime point
of view
\begin{eqnarray}
X^0 & = & t \nonumber\\
X^1 & = & r(t) \cos  \theta \nonumber\\
X^2 & = & r(t) \sin  \theta \cos  \phi
\label{S3inducedmetric}
\\
X^3 & = & r(t) \sin  \theta \sin  \phi \nonumber\\
X^4 & = & \chi \nonumber\\
F_{t \chi} & = & E(t) \nonumber 
\end{eqnarray}
\begin{eqnarray}
ds^2=
(-1+\dot{r}^2)dt^2 + d\chi^2 + r^2(t) [d\theta^2 +
\sin ^2 \theta d \phi^2],
\end{eqnarray}
where we parametrize the worldvolume of the S3-brane by
$(t,\theta,\phi,\chi)$.  At any given moment, the S-brane
worldvolume is a cylinder, $R\times S^2$. The open string metric
and its inverse are
\begin{eqnarray}
&& (g+F)_{ab}=
\bordermatrix{
&t             & \chi  & \theta &  \phi    \cr
&-1+\dot{r}^2  & E(t)  & 0      & 0     \cr
&-E(t)         & 1     & 0      & 0   \cr
&0             & 0     & r^2    & 0  \cr
&0             &  0    &  0     & r^2 \sin ^2\theta \cr
},
\\
&& (g+F)^{ab}= \frac{1}{-1\!+\!\dot{r}^2\! +\! E^2}
\bordermatrix{
&t      & \chi        \cr
&1      & -E(t)       \cr
&E(t)   & -1\! +\!\dot{r}^2 \cr
} \ .
\end{eqnarray}
The Lagrangian for this S-brane is (up to a normalization constant
for the S-brane tension)
\begin{equation}
\sqrt{\det(g+F)}= r^2 \sin  \theta \sqrt{-1+\dot{r}^2+E^2}
\end{equation}
and the equation of motions for the embedding are
\begin{equation}
\partial_a \left( \sqrt{\det(g+F)} (g+F)^{ab} \partial_b X^M\right)=0,
\end{equation}
where $M=0,\cdots,4$.
There are only two distinct equations of motion for this system
(the gauge field equations of motion can also be checked), the
first of which is
\begin{equation}
\partial_a
\left( r^2 \sin \theta \sqrt{-1+\dot{r}^2+E^2} (g+F)^{ab}
\partial_b t\right)= 0
\end{equation}
while the second equation of motion is
\begin{eqnarray}
\partial_a \left[
r^2 \sin \theta \sqrt{-1+\dot{r}^2 +E^2} (g+F)^{a b}
\partial_b (r \cos  \theta)\right] = 0.
\end{eqnarray}
We use the first equation of motion to simplify the derivative
term in the second equation of motion and then rearrange
terms slightly, to obtain
\begin{eqnarray}
\partial_t \! \left(
\frac{r^2}{\sqrt{-1+\dot{r}^2 +E^2}} \right)=0,
\quad
r \ddot{r} + 2 (1-\dot{r}^2 - E^2) = 0.
\end{eqnarray}
Finally, substituting the second equation into the first, we get
the differential equation for the radius
\begin{equation}
\partial_t
\left(\frac{r^{3/2}}{\sqrt{\ddot{r}}}\right)=0 \ \quad
\Leftrightarrow  \quad
\ddot{r}= A
r^3 \label{diffeqS3brane}
\end{equation}
which has a solution describing the confinement of electric flux
\begin{equation}
r=\frac{c}{t}, \hspace{.3in} E=1.
\label{confof}
\end{equation}
The electric field is always constant and takes the critical
value, while the radius of this flux tube shrinks to zero at
$t=\infty$. The electric field is necessarily constant since
there are no magnetic fields; a changing electric field would
necessarily also produce a magnetic field.  Although this
solution only exists for $t>0$, this does not mean that the
dynamics on the non-BPS mother brane is trivial for $t<0$. Before
$t=0$ it is still possible to have flux on the non-BPS mother
brane and yet no $T=0$ regions. The key point is that the S-brane
is only defined where the tachyon value is zero and so captures
partial knowledge of the full tachyon configuration and flux.
Yet, at the same time there is no violation of fundamental string
charge from the S-brane viewpoint.\footnote{For this solution
(\ref{confof}) the total fundamental string number is $4\pi c$.
See Eq.~(\ref{ften}).} This S-brane comes in from spatial
infinity and brings in charge through the gauge fields on its
worldvolume. For charge conservation we do not have to have time
reversal S-brane solutions which would correspond to including a
mirror copy of the above solution describing an expanding
worldvolume. We point out however, that the expanding string
solution is interesting in its own right and is possibly related
to instabilities due to critical electric fields and possibly the
Hagedorn temperature. Further discussion on why this solution
represents a fundamental string at late times is given in
Ref.~\cite{Sbraneaction}.

These spike solutions correspond to inhomogeneous tachyon
configurations which spontaneously localize into lower
dimensional systems.  An example of such a solution was found by
Sen in Ref.~\cite{Sentimeevole}.

\subsection{Discussion on confinement}
\label{doc}

In Ref.~\cite{Sbraneaction} and in the previous subsection, we
have seen S-brane solutions describing the decay of an unstable
D-brane into fundamental strings. A peculiar feature of these
solutions is that eventually the electric flux becomes
concentrated around the S-brane remnant where $T = 0$. Is this a
generic phenomenon corresponding to the confinement\footnote{See
Ref.~\cite{Kuroki} for a discussion on the dielectric effect on
classical confinement of fluxes, and also Ref.~\cite{Yi, GHY1}
for the confinement on branes.} of fundamental strings? In this
subsection, we will discuss how the S-brane configuration is
related to confinement in a tachyon system by showing that it is
the lowest energy configuration for fixed electric flux.
Furthermore, the magnetic field is also shown to be classically
confined, which is consistent with the S-brane solution of
D-string formation presented in Section \ref{dfs}.

The main idea is that as an unstable D-brane decays, the tachyon
condenses $T\rightarrow \infty$ almost everywhere except at the
location of the S-brane remnant where $T=0$. We wish to show that
the electric flux will concentrate around the region $T = 0$.

Take an unstable D2-brane for simplicity. To begin, let us first
consider homogeneous configurations with electric field $F_{01} =
E$. The Lagrangian density is of the form
\begin{equation} \label{L0}
{\cal L} = -\sqrt{1-E^2}\Lag(T, z),
\end{equation}
where
\begin{equation}
z = -\frac{\dot{T}^2}{1-E^2},
\quad E = \dot{A},
\end{equation}
and this Lagrangian is valid for $0 \leq E^2 <1$.
The conjugate variables of $T$ and $A$ are
\begin{eqnarray}
P &=& \frac{\del \cal L}{\del\dot{T}} =
\frac{1}{\sqrt{1-E^2}}\frac{\del\Lag}{\del z}\dot{T}, \\
D &=& \frac{\del \cal L}{\del E} = \frac{E}{\sqrt{1-E^2}}
\left(\Lag -2z \frac{\del\Lag}{\del z}\right) \label{Pi}
\end{eqnarray}
so the Hamiltonian density is
\begin{equation} \label{H0}
{\cal H} = P\dot{T} + D E - {\cal L} =
\frac{1}{\sqrt{1-E^2}}
\left(\Lag - 2z \frac{\del\Lag}{\del z}\right).
\end{equation}
As long as
$E \neq 0$, we have the simpler expression \cite{GHY2}
\begin{equation}
\label{H1}
{\cal H} =
\frac{D}{E}.
\end{equation}

Now consider those configurations which can be approximated by a
homogeneous region for $|x_2| < l/2$, and a different homogeneous
region when $|x_2| > l/2$.  For our purposes the two regions will
correspond to the S-brane region $T = 0$, and the tachyon
condensation region $T\rightarrow \infty$. When the D2-brane decays,
some energy will be dissipated or radiated away but the electric flux
\begin{equation}
\Phi = \int dx^2 D
\end{equation}
will be preserved.
The final state of the process should be
the most energy-efficient configuration for a given flux.

According to (\ref{H0}),
the energy in the region of tachyon condensation
can be arbitrarily close to zero.
As an example, for the effective theory with
$\tilde{\cal L} = V(T) f(z)$,
where $V(T)\rightarrow 0$ as $T \rightarrow \infty$,
we can set $T \rightarrow \infty$ and $\dot{T} \rightarrow 0$
such that ${\cal H} = 0$.
It follows from (\ref{Pi}) that
$D = 0$ in the condensate region as long as $E < 1$. Although
there is electric field everywhere on the non-BPS brane, the
flux is only non-zero in the S-brane region
\begin{equation} \label{constraint}
l D = \Phi,
\end{equation}
where
$D$
is the electric flux density for $|x_2| < l/2$.
The total energy is
\begin{equation}
H = l {\cal H} =
l\frac{D}{E}
= \frac{\Phi}{E},
\end{equation}
where we used (\ref{H1}).
Since $\Phi$ is a given fixed number, the energy $H$ is minimized by
maximizing $E$.
We conclude that the minimal energy state has
\begin{equation} \label{E1}
E \rightarrow 1
\end{equation}
around the S-brane,
and so the energy is from pure flux $H = \Phi$, that is,
the total energy is the same as the energy
due to the tension of the fundamental strings.
Finally, due to Eq.~(\ref{Pi}), in the limit where the electric
field goes to the critical value,
$D \rightarrow \infty$,
and so the width of the S-brane region
with nonzero electric flux shrinks to zero
\begin{equation}
l = \frac{\Phi}{D} \rightarrow 0.
\end{equation}
We have therefore shown that the electric flux is confined
to the infinitesimal region around $T = 0$.

We hope that the analysis above captures
the physical reason for confinement
in the low energy limit and with the present result one can show that the
confined flux behaves as a fundamental string governed by a Nambu-Goto
action following the argument given in Refs.~\cite{GHY1,GHY2}.
In the above discussion however we ignored the transition
interpolating the two homogeneous regions.
When the transition region is taken into account,
it might happen that the confinement profile
has an optimal width at some characteristic scale.

Is there confinement for the magnetic flux as well?
Since S-duality interchanges fundamental strings
with D-strings, we expect the answer to be yes.
We will study the consequences of S-duality for S-branes in the next
section, while here we will continue with a direct analysis of the
tachyon system.
It is well known that a magnetic field on a BPS D$p$-brane gives a
density of lower-dimensional BPS D$(p-2)$-branes on the mother
D-brane. Naively, however, a magnetic field on the non-BPS brane
does not give any lower dimensional BPS D-brane charge. The
effect of the magnetic field appears only on the tachyon defects.
For example on a non-BPS D3-brane, a tachyon kink is equivalent
with a BPS D2-brane. Suppose that we have a magnetic field on the
original non-BPS brane along the kink. Then this induces BPS
D0-brane charge only on the D2-brane, while apart from the kink
no charge is induced though the magnetic field is present all over the
non-BPS D3-brane worldvolume.

Keeping the above charge conservation in mind,
let us try the same confinement argument to tackle this problem.
The analogue of Eq.~(\ref{L0}) is
\begin{equation}
{\cal L} = -\sqrt{1+B^2}\Lag(T, z),
\end{equation}
where $z = -\frac{1}{2}\dot{T}^2$,
and the analogue of Eq.~(\ref{H0}) is
\begin{equation}
{\cal H} =
\sqrt{1+B^2}
\left(\Lag - 2z \frac{\del\Lag}{\del z}\right).
\end{equation}
As in the case of electric flux, we consider a homogeneous
S-brane region\footnote{Although a homogeneous tachyon profile
$T=0$ will not help to give the lower dimensional RR charge
because the RR coupling on the non-BPS brane is proportional to
$dT \wedge F$ while $dT$ vanishes, we believe that the argument
here captures an important feature of confinement. } of width $l$
and a tachyon condensation region.
Let the magnetic fields in the two regions be $B_0$ and $B_1$.
The energy in the condensed region can be
minimized to zero by assigning $T \rightarrow \infty$
and $\dot{T} = 0$.
The total energy is
\begin{equation}
H = l {\cal H} = C l\sqrt{1+B_0^2},
\end{equation}
where $C$ is a constant independent of $B_0$ and $l$.
This energy $H$ is to be minimized with the constraint
that the total flux only on the S-brane region is conserved
(or to assume that $B_1 = 0$),
that is
\begin{equation}
\Phi_0 = lB_0 = \mbox{fixed}.
\end{equation}
Using the same arguments as before, we see that $H$ is minimized
for $l = 0$ (and also $B_0 \rightarrow \infty$), which shows the
confinement of the lower dimensional RR charge.

We will see in Section \ref{dfs} that in fact one can construct
an S3-brane spike solution which represents the formation of
$(p,q)$ strings.
The argument for confinement of electric and
magnetic fields we have presented here is therefore consistent with our
interpretation of the spike solutions.

\section{D-branes from S-branes}
\label{dfs}

In the previous section, we reviewed the formation of fundamental
strings from S-branes and showed how confinement of electric flux
can be a strong coupling but classical process. We found also
that magnetic flux on a non-BPS brane is confined, which was
expected due to the electric-magnetic duality in string theory.
Confinement of magnetic fields should occur in any theory with
electric-magnetic duality with confined electric flux bundles. In
string theory the electric fluxes act as fundamental strings
while confined magnetic fluxes act as branes; D-strings will be
the focus of our attention. In this section we show how an
S3-brane can realize the dynamical formation of $(p,q)$ string
bound states and D-strings, and so in a similar vein this will
demonstrate that magnetic fields also confine. Magnetic fields
can have an effect on tachyon dynamics.

\begin{figure}[bhtp]
\begin{center}
\epsfxsize=6.5in\leavevmode\epsfbox{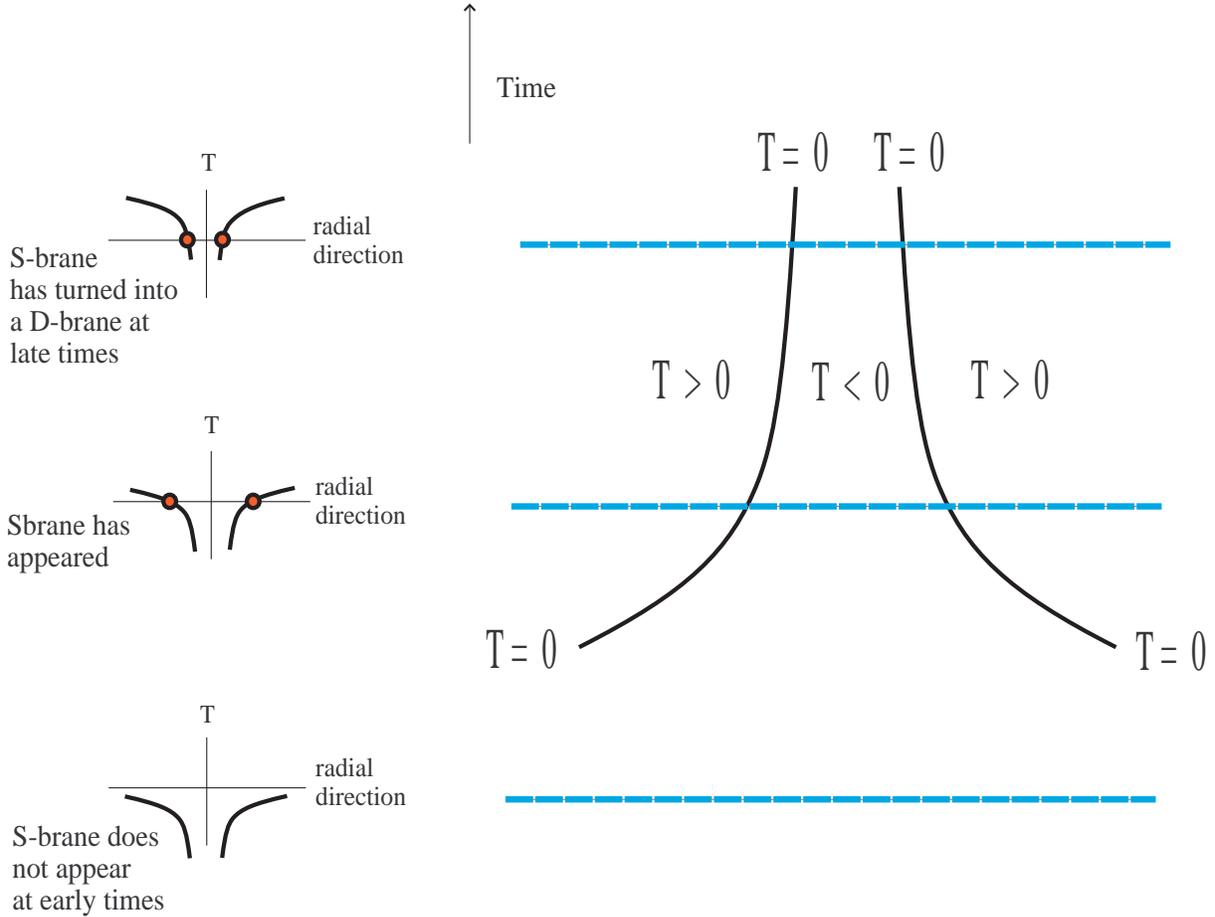}
\caption{The figure on the right is a schematic cross section of
tachyon values on the non-BPS brane which gives rise to a decaying
S-brane.  To the left we have included snapshots of the tachyon
values at specific times.  At early times the tachyon
configuration is changing but an S-brane has not appeared.  The
S-brane then appears, coming in from infinity, and then slows
down to metamorphose into a D-brane. The tachyon configuration is
not a kink or lump but more like an infinite well. Time dependent
kinks do not necessarily leave spatial kink remnants. Related
discussion can be found in Section \ref{bsas} and \ref{sadi}.}
\label{tachyonvalues}
\end{center}
\end{figure}

Another motivation for searching for these solutions is the fact
that, as opposed to fundamental strings, it is already known that
D-branes can be described in the context of tachyon condensation.
If we can discuss D-branes formation using S-branes then the
related tachyon solutions should be easier to obtain. (An
understanding of tachyon solutions would also help to explain how
to construct closed strings from an open string picture.)  A
schematic cross section of expected tachyon values is shown in
Fig.~\ref{tachyonvalues}. From this illustration we see that
while the S-brane region ($T=0$) seems to appear ``out of
nowhere'' and therefore seems to violate causality, from the
tachyon picture there is in fact no difficulty.  Before the
S-brane appears, the tachyon field is simply evolving with no
$T=0$ regions.
Also at very early times, the entire spacetime is filled with only
one of the vacua and so it is impossible to consider stable lower
dimensional defects.
When
the tachyon has evolved closer to the second vacuum
at late times, it is possible to interpret the $T=0$ regions as
physical objects. By the time we can interpret the S-brane as a
standard localized defect, it has already slowed down to less than
the speed of light.

\subsection{Tachyon solutions with homogeneous
            electric/magnetic fields}
\label{tswhemf}
\setcounter{footnote}{0}

Before turning to the formation of $(p,q)$ strings, we first
consider homogeneous tachyon solutions with magnetic fields in
analogy to the electric case in Ref.~\cite{GHY2}.

To better understand the tachyon condensate, it has been proposed
\cite{GHY2} that in the effective action description of non-BPS
branes
\begin{equation}
L=V(T)\sqrt{-\textup{det}(\eta+F)} {\cal{F}}(z),
\end{equation}
\begin{equation}
z \equiv ((\eta+F)^{-1})^{\mu \nu}
\partial_\mu T \partial_\nu T =
[(\eta-F\eta^{-1}F)^{-1}]^{\mu\nu}
\del_{\mu}T \del_{\nu}T,
\label{z}
\end{equation}
not only does the potential go to zero but that the kinetic energy
contribution of the tachyon also vanishes
\begin{equation}
{\cal F}(z)=0  \ \ \Leftrightarrow \ \ z=-1 \
\label{critconstraint}
\end{equation}
after tachyon condensation. For uniform electric fields and
tachyon fields this leads to a constraint
\begin{equation}
\dot{T}^2+E^2=1
\end{equation}
which governs the tachyon system near the bottom of the
tachyon potential.\footnote{
In Ref.~\cite{GHY2}, this condition $z=-1$ comes from
requiring that
$D$
and $H$ be preserved while $V(T)\rightarrow 0$
for a homogeneous background.}
One motivation for searching for such a constraint is that
it should help to describe confinement of electric flux on a non-BPS
brane, and it was shown that this constraint leads to a Carrol
limit for the propagating degrees of freedom on the brane.  The
effect of the Carrol limit is to make the condensate a fluid of
electric strings.

It is straightforward to extend the above analysis to include
magnetic fields as well as electric fields.
For simplicity we explicitly work out the $2+1$ dimensional case
but all other cases can be treated in the same manner.
Similar discussion
has also recently appeared in Ref.~\cite{KKKK-EMfluids}.

When the fields are all spatially homogeneous the open string
metric is
\begin{eqnarray}
(g+F)_{ab}= \bordermatrix{ &t              & x      & y \cr
&-1             & E_x    & E_y    \cr &-E_x           & 1      &
B      \cr &-E_y           & -B     & 1      \cr }
\end{eqnarray}
and to calculate the constraint we only need the $G^{tt}$
component of the inverse of this matrix.  A simple calculation
shows that the constraint $z=-1$ becomes
\begin{equation}
\dot{T}^2 + \frac{E^2}{1+B^2} =1 \ .
\label{TEB}
\end{equation}
There is no obvious duality between electric and magnetic fields
since the tachyon scalar field breaks the worldvolume Lorentz
invariance.
The effect of the magnetic field is to increase the critical
electric field,
and if we take $\dot{T}=0$ then we
reduce to the simple Lorentz invariant condition
\begin{equation}
E^2-B^2=1
\ . \label{ebconst}
\end{equation}

The role played by electric and magnetic fields is interesting
and we make the following observations.  First, a critical
electric field will stop the tachyon from rolling near the end of
the tachyon condensation process.  Second, it has been shown that
a D-\Dbar\ pair with critical electric field is supersymmetric
\cite{supertube}.
Even though these results were derived in different contexts,
there is an overlap in the way a critical electric field on branes
removes tachyon dynamics and one wonders if there are further
connections. For example perhaps the reason why the tachyon
ceases to roll in the presence of the electric field is also due
to supersymmetry.  In general we should be able to see regions of
supersymmetry develop during the tachyon condensation process,
where $\dot{T}=0$ and these regions could have interpretations as
various lower dimensional supersymmetric objects.  We further
point out the existence of supersymmetric D-\Dbar\ configurations
which are distinct \cite{MyersWinters-DDbarEMfields} from the
critical electric field case. These solutions should also appear
as end products of tachyon condensation and be related to
different constraints on the tachyon Lagrangian.

As we have just observed in $2+1$ dimensions, if there are no
electric fields, then there is apparently no effect due to the
magnetic field near the tachyon minimum.  For higher dimensions,
it is clear that if we follow similar steps, the homogeneous
magnetic field by itself does not effect tachyon dynamics.  One
way to understand why the magnetic field does not change the
rolling tachyon condition is that a constant magnetic field on a
non-BPS D$p$-brane can be understood as a bound state of a non-BPS
D$p$-brane and non-BPS D$(p\!-\!2)$-branes. Both of these have a
rolling tachyon ($\dot{T}=1$), so the resultant bound state also
has the rolling tachyon. Constant magnetic fields in this
situation are not capable of generating stable lower dimensional
objects. On the other hand, more complicated configurations with
magnetic fields can create lower dimensional branes as we will
see in the next subsection.

Finally, let us obtain the
results of Eq.~(\ref{critconstraint}) from the worldsheet
point of view.  An open string on the
D-brane has opposite charges at its endpoints. In a constant
electric field background, the charges are pulled in opposite
directions, with the electrostatic force in competition with the
tension. When we stretch a string in an electric field which is
strong enough ($ E = \pm 1 $), the increase in energy due to
tension is compensated by the decrease in electric potential
energy. The strings can have infinite length with vanishing
energy. It appears as if the strings have no tension, resembling
a collection of particles or dust. We propose to interpret this
situation as tachyon condensation, or the point at which the
D-brane vanishes.

Consider an open string with the worldsheet action
\begin{eqnarray}
\label{S}
S& =&\int\!d^2\sigma\; \left[\frac{1}{2}(\dot{X}^2-X'{}^2
+F_{\mu\nu}\dot{X}^{\mu}X'{}^{\nu})+X'{}^{\mu}\del_{\mu}\Phi(X)\right]
\nonumber \\
&=&\int\! d^2\sigma\;
\frac{1}{2}(\dot{X}^2-X'{}^2) +\int\! d\tau \left(
-\frac{1}{2}F_{\mu\nu}X^{\mu}\dot{X}^{\nu}+\Phi(X)\right).
\end{eqnarray}
The spacetime momentum densities are
\be
\label{P}
P_{\mu} = \dot{X}_{\mu}+F_{\mu\nu}X'{}^{\nu}.
\end{equation}
The equation of motion is
\begin{equation}
\ddot{X}^{\mu}-X''{}^{\mu} = 0,
\end{equation}
and the
boundary condition is
\begin{equation}
\label{BC}
X'_{\mu}+F_{\mu\nu}\dot{X}^{\nu}=\del_{\mu}\Phi(X),
\end{equation}
at the string endpoints $\sigma=0,\pi$.

We don't consider oscillation modes, so we impose the above
boundary condition on the whole string. From (\ref{P}) and
(\ref{BC}), we obtain the relation
\begin{equation} (\delta^{\mu}_{\nu}-F^{\mu\kappa}F_{\kappa\nu})X'{}^{\nu}
=-F^{\mu\nu}P_{\nu}+\del^{\mu}\Phi(X).
\end{equation}
{}From this equation we see that there are solutions with
arbitrarily large $X'$ and $P_{\mu}=0$ (that is, arbitrarily long
strings at no cost in energy or momentum) if either
\begin{equation} \label{1+F} \det(1-F^2) = \det(1+F)\det(1-F) =
(\det(1+F))^2 = 0,
\end{equation} or \begin{equation} \label{delPhi} \del_{\mu}\Phi =
\infty. \end{equation}

The first condition (\ref{1+F}) agrees with (\ref{TEB})
when $\dot{T} = 0$.
The second condition (\ref{delPhi}) agrees with
the final state of the rolling tachyon solution of Sen
\begin{equation}
\Phi \propto e^{X^0}.
\end{equation}
It can be related to the desired condition for $T$ via a change
of variable such as
\begin{equation}
\Phi = \frac{T}{\sqrt{1+z}},
\end{equation}
where $z$ is defined in (\ref{z}).
The condition (\ref{delPhi}) is now
\begin{equation} z =
-1. \end{equation}

\subsection{S3-branes with electric and magnetic fields}
\label{s3bwmf}

Let us proceed to construct a solution of the S3-brane action
which represents a formation of a $(p,q)$ string bound state. The
ansatz is identical to the one in the previous section,
Eq.~(\ref{S3inducedmetric}), except that we also include an
additional magnetic field
\begin{eqnarray}
X^0 & = & t \nonumber \\
X^1 & = & r(t) \cos  \theta \nonumber \\
X^2 & = & r(t) \sin  \theta \cos  \phi \nonumber \\
X^3 & = & r(t) \sin  \theta \sin  \phi \label{paraeb}\\
X^4 & = & \chi \nonumber \\
F_{t \chi} & = & E(t)
\nonumber \\
\quad F_{\theta \phi} & = & b \sin \theta \ . \nonumber
\end{eqnarray}
The open string metric and its inverse are just direct products of
the example we gave before and
\begin{equation}
(g+F)_{ab}= \bordermatrix{ &\theta & \phi \cr &r^2 \sin ^2 \theta
& b \sin \theta  \cr & - b \sin \theta & r^2 } \nonumber
\end{equation}
\begin{equation}
(g+F)^{ab}= \frac{1}{r^4 \sin ^2 \theta(1+\frac{b^2}{r^4})}
\bordermatrix{ &\theta             & \phi            \cr &r^2
\sin ^2 \theta & -b \sin \theta  \cr &b \sin \theta      & r^2 }
\end{equation}
so the action is proportional to
\begin{equation}
\sqrt{\det(g+F)}=  r^2 \sin \theta \sqrt{(-1+\dot{r}^2
+E^2)\left(1+\frac{b^2}{r^4}\right)} \ .
\end{equation}

We first examine the equation of motion of the embedding
coordinate
\begin{equation}
\partial_t
\left[ r^2 \sin \theta \sqrt{(-1+\dot{r}^2+E^2)
\left(1+\frac{b^2}{r^4}\right)} (g+F)^{tt} \partial_t t\right]
= 0
\label{firsteq1}
\end{equation}
and try a solution of the form
\begin{equation}
r=\frac{c_{\rm d}}{t} ,\hspace{.3in} E=\textrm{const.} \ ,
\end{equation}
where $c_{\rm d}$ is a constant parameter.
This ansatz gives a solution as long as we satisfy the relation
\begin{equation}
E^2-\frac{b^2}{c_{\rm d}^2}=1 \label{S3constraint}
\end{equation}
which is consistent with the constraint in
Eq.~(\ref{critconstraint}) since on the S-brane worldvolume
$\dot{T}=0$. It is straightforward to check that the other
equations of motion such as
\begin{equation}
\partial_a \left[ r^2 \sin\theta \sqrt{(-1+\dot{r}^2+E^2)
\left(1+\frac{b^2}{r^4}\right)} (g+F)^{ab} \partial_b (r
\cos\theta) \right] =0
\end{equation}
and
\begin{equation}
\partial_a \left[ r^2 \sin\theta \sqrt{(-1+\dot{r}^2+E^2)
\left(1+\frac{b^2}{r^4}\right)} (g+F)^{ab} \right]
=0
\end{equation}
are also satisfied.  The field strength $F_{\theta\phi}$
generates a magnetic field along $\chi$ and parallel to the
electric field.  This S-brane is an electric-magnetic flux tube
confining into a $1+1$ dimensional remnant.  At late times this
S-brane becomes a $(p,q)$ string bound state.   The existence of
these additional solutions should be expected due to S-duality on
the S3-brane as we will explain in the following subsection.

We note that these solutions have real $F_{\theta \phi}$ as long
as the electric field is greater than the critical value due to
Eq.~(\ref{S3constraint}). Although the appearance of large
electric fields is unusual on D-branes, they appear quite
naturally on S-branes and large electric fields do not lead to
imaginary S-brane actions. We will see in the following section
how large electric fields show up on S-branes by examining the
tachyon solutions on the non-BPS mother branes.

\subsection{S-duality for S3-branes}
\label{sis}

For the purposes of this subsection, the following parametrization
\begin{eqnarray}
 X^0 & =& X^0 (x^1,x^2,x^3) \nonumber \\
 X^1 & =& x^1 \nonumber \\
 X^2 & =& x^2 \nonumber \\
 X^3 & =& x^3 \\
 X^4 & =& \chi \nonumber \\
 F_{a\chi} & = & \p_b A_\chi(x^1,x^2,x^3) \nonumber \\
 F_{ab} & = & F_{ab}(x^1,x^2,x^3) \nonumber
\end{eqnarray}
turns out to be useful to see the duality transformations, where
$a,b=1,2,3$.  The worldvolume of the S3-brane is now parameterized
by $(x^1,x^2,x^3,\chi)$
as in Ref.~\cite{Sbraneaction}. In the above parametrization we
have assumed that all the fields are independent of $\chi$ just
like in our explicit S-brane solutions.
We follow Ref.~\cite{Gibbons} in deriving the extended duality
symmetry. This will help clarify how the $(p,q)$ string formation
solutions are related to the F-string formation solution in
Section \ref{sfs}, and suggests other ``non-BPS'' throat-type
solutions.

The S3-brane Lagrangian in this coordinate choice is written as
\cite{Sbraneaction}
\begin{eqnarray}
L& =& \sqrt{\det (\delta_{ij} - \p_i X^0 \p_j X^0 + F_{ij})}
\\
&=&
\left[1-(\p_a X^0)^2 + (F_{a\chi})^2 + (F_{ab})^2/4 +
(\p_a X^0F_{a\chi})^2
\right.
\nonumber
\\
& & \hspace{30mm} \left. - (\p_a X^0)^2(F_{b\chi})^2 -
(\epsilon^{abc}F_{bc}\p_a X^0)^2/4 +
(\epsilon^{abc}F_{bc}F_{a\chi})^2/4 \right]^{1/2} \ . \nonumber
\end{eqnarray}
We omit the overall constant factor $S_0$ in the S-brane
Lagrangian. We next introduce the Lagrange multiplier field
$\phi_B$ for the Bianchi identity of $F_{ab}$ as
\begin{eqnarray}
\Delta L = (\phi_B/2)[\epsilon_{abc}\p_aF_{bc}].
\label{multerm}
\end{eqnarray}
With this multiplier term we can regard $F_{ab}$ as fundamental
fields
and integrate out the field strength $F_{ab}$.  The
final form of the Lagrangian is simply
\begin{eqnarray}
 L+ \Delta L
= \sqrt{\det\left(\eta^{RS}+ \nabla\Phi^R\cdot \nabla\Phi^S\right)}
\label{laggen}
\end{eqnarray}
where $\Phi^R = (X^0, \phi_B, A_\chi)$ and the metric in the
virtual transverse space parametrized by $\Phi^R$ is
$\eta^{RS}={\rm diag}(-1,-1,1)$. The Lagrangian shows that the
whole duality symmetry is $SO(2,1)$.  It is interesting that the
subgroup of the duality symmetry which rotates the electric and
magnetic fields $A_\chi$ and $\phi_B$, in other words $F_{a\chi}$
and $\epsilon_{abc}F_{bc}$ and so this should be S-duality, is in
this case $SO(1,1)$.  This duality is more like a Lorentz boost
between electric and magnetic fields than the usual duality
rotations.  We will explain how to obtain the more usual duality
rotations in the next subsection.

If $X^0$ and $A_\chi$ are turned on,
the duality group becomes $SO(1,1)$ whose fixed point is the spike
solution
\begin{eqnarray}
 X^0 = A_\chi = \frac{c}{r}
\end{eqnarray}
which represents the formation of a fundamental string.
If we also turn on $\phi_B$, we obtain the spike solution
representing the formation of a $(p,q)$ string at late time
\begin{eqnarray}
 X^0 = \frac{A_\chi}{\cosh \alpha} = \frac{\phi_B}{\sinh\alpha}
= \frac{c_{\rm d}}{r}
\end{eqnarray}
which we provided in the previous subsection from an alternate
viewpoint, Eq.~(\ref{ebconst}).  The relationship between these two
parametrizations is
\begin{eqnarray}
 E = \cosh \alpha, \quad  b=c_{\rm d} \sinh \alpha.
\end{eqnarray}
The fundamental string charge and the D-string charge
are $4\pi c_{\rm d}\cosh\alpha$ and $4\pi c_{\rm d}\sinh\alpha$,
respectively.

The duality group $SO(1,1)$ above is only
a subgroup of the full S-duality symmetry group $SL(2)$.
(It becomes $SL(2,Z)$ upon charge quantization.)
Here we started off with an S-brane solution which decayed into
fundamental strings $(n,0)$.  The $SO(1,1)$ symmetry connects it
to $(p,q)$ strings with $p>q$, but we are still missing all other
$(p,q)$ strings with $p<q$. We will discuss how to obtain these
other cases in the next subsection.

\subsection{Magnetic S-branes from M-theory}
\label{msfm}

Since D-branes can be realized as defects on the non-BPS brane
worldvolume, one is tempted to try to find the S3-brane spike
solution decaying into just D-strings. However, the condition in
Eq.~(\ref{S3constraint}) implies that if there are no fundamental
strings ($E=0$), there is no solution with real magnetic fields;
one can prove this from the equations of motion
Eq.~(\ref{firsteq1}) and assuming only rotational symmetry.
In the context of tachyon condensation of non-BPS branes, the
magnetic field on any S-brane induced from the field strength on
the corresponding non-BPS brane should be real. Magnetic
solutions do exist, however, if we allow for imaginary field
strengths.  It is possible to investigate the implications for
allowing imaginary field strength solutions.
Imaginary field strengths have been
noted to potentially arise in time dependent systems \cite{Hull}
and it remains to be seen whether they will play a physical role
in a theory.

However, instead of introducing imaginary field strengths we will
find a way to mimic their behavior with real magnetic fields and
so avoid the constraint of Eq.~(\ref{S3constraint}).  The key
point will be to consider M-theory effects by dualizing the
scalar field from the M-theory circle to a gauge field.  This
dualized gauge field will not be induced from the non-BPS brane
but from M-theory.

Earlier in Section 2, we discussed a generalized S-brane action
for spacetime vortices in Eq.~(\ref{vortexaction}). The main
difference was that this generalized action included fluctuations
of a transverse scalar along a spatial direction. Up to now we
have not used this scalar, however we will now use this to solve
the riddle of D-string generation from an S3-brane.  The idea is
to consider M-theory compactified on two circles, one the
M-theory circle which reduces us to type IIA and another to take
us from type IIA to type IIB string theory.

In this case we can begin with an M5-$\overline{M5}$ pair and look
for a codimension three generalized vortex solution representing a
spacelike M2-brane. This should be present just by generalizing
the argument in Ref.~\cite{Yi} where an M2-brane is realized as a
topological soliton in M5-$\overline{M5}$. The spacelike M2-brane
Lagrangian of this spacetime vortex is
\begin{equation}
L=\sqrt{\textup{det} (\delta_{ij}  - \p_i X^0 \p_j X^0
+ \p_i X^4 \p_j X^4
+ \p_i X^{10} \p_j X^{10} )}, \hspace{.3in} i,j=1,2,3
\label{spvortex}
\end{equation}
where the spatial transverse direction is along the M-theory
circle $X^{10}$. Let us
dualize the scalar $X^{10}$ into a gauge field with field strength
$\tilde{F}$.
We perform the dualization by adding the Lagrange
multiplier term
\begin{equation}
\Delta L = \frac12 X^{10} \epsilon_{ijk} \p_i \tilde{F}_{jk}
\end{equation}
and then integrating out $X^{10}$.  The final form of
the Lagrangian is
\begin{equation}
L + \Delta L=\sqrt{\textup{det} (\delta_{ij} - \p_i X^0 \p_j X^0 +
\p_i X^4 \p_j X^4 + i \tilde{F}_{ij} )} \label{imaginaryFaction}
\end{equation}
where the factor of ``$i$'' now accompanies the dual field
strength!  This factor does not need to be added into the
Lagrange multiplier term but instead is a direct consequence of
the Euclidean nature of the S-brane action. If the scalar $X^4$
is trivial as in the present situation $X^4=\chi$ of Sections
\ref{sfs} and \ref{dfs}, this S2-brane action in type IIA can be
regarded as an S3-brane action in type IIB theory. In this action
we can now solve for a purely magnetic S3-brane solution as in
Section \ref{s3bwmf} but now with real field strength. We
emphasize that the field strength is real and the factor of
``$i$'' does not
effect the hermiticity of the action.  One might ask if the
S3-brane can be constructed directly from an unstable 4-brane
object. It is possible that this S3-brane construction can be
studied on the S-dual of the non-BPS D4-brane which has also been
called an NS4-brane \cite{Kleban}.

This S3-brane decays into a one dimensional remnant with magnetic
field, so our expectation should be that this is a D-string.  Let
us obtain the explicit solution and see how the D-string tension
is reproduced.  We begin with the action for this magnetic
S3-brane written in the spacetime point of view with the
parametrization (\ref{paraeb}),
\begin{equation} L=r^2 \sin  \theta
\sqrt{-1+\dot{r}^2+B^2-B^2\dot{r}^2} , \hspace{.3in} B \equiv
\frac{\tilde{F}_{\theta \phi}}{r^2\sin\theta} \ .
\end{equation}
The solution for this decaying S3-brane is
\begin{eqnarray}
 r = \frac{c_{\rm m}}{t}, \quad B = \frac{c_{\rm m}}{r^2},
\end{eqnarray}
where we take $c_{\rm m}$ to be positive.
We calculate the conjugate momenta and Hamiltonian
\begin{eqnarray}
P_r & = & r^2 \sin \theta
\frac{\dot{r}[1-B^2]}{\sqrt{-1+\dot{r}^2+B^2-B^2 \dot{r}^2}} \ , \\
H\equiv \int\! d\chi d\theta d\phi \left[ P_r \dot{r}- L \right]
 & =& \int\! d\chi d\theta d\phi \;
r^2 \sin \theta \frac{-1+B^2}{\sqrt{-1+\dot{r}^2+B^2-B^2
\dot{r}^2}} \ .
\end{eqnarray}
At late times $\dot{r}=0$ and $B$ is large so
in this limit the Hamiltonian has the simple form
\begin{equation}
H= \int\! d \chi \int_{S^2} d\theta  d\phi\; B r^2 \sin\theta
= 4\pi c_{\rm m} \int\! d\chi.
\end{equation}
At this stage we recall that in the above analysis we omitted the
overall factor of the S3-brane tension, and also that the
parameter $c_{\rm m}$ should be subject to Dirac quantization
condition. It is naturally expected that the S3-brane tension is
given by the D3-brane tension, $T_{\rm D3} = 1/2\pi g_{\rm s}$ in
our convention $2\pi\alpha'=1$.  Now what about the Dirac
quantization condition?  Let us compare this magnetic S-brane
with the electric case in the previous section. A straightforward
calculation shows that the energy there is given by the same
expression
\begin{eqnarray}
 H_{\mbox{\scriptsize electric S3}} = 4\pi c\int\! d\chi
\label{ften}
\end{eqnarray}
where again the overall tension $T_{\rm D3}$ is omitted, and $c$
is the parameter appearing in the solution (\ref{confof}). Now
the Dirac quantization condition is
\begin{eqnarray}
 4\pi c\cdot 4\pi c_{\rm m} = \frac{2\pi n}{T_{\rm D3}}
\label{Dcon}
\end{eqnarray}
where $n$ is an integer and the factors of $4\pi$ come from
integrating over the $S^2$ angular directions of the S-brane
worldvolume. The factor $T_{\rm D3}$ appears here since this
factor appears in the action and so the right hand side is
proportional to the string coupling constant $g_{\rm s}$.

Let us see how this condition works. What we are doing is a
generalization of Ref.~\cite{CallanMaldacena}. Suppose that
Eq.~(\ref{ften}) gives the correct tension of a fundamental string,
\begin{eqnarray}
 4\pi c T_{\rm D3}= T_{\rm F1}
\end{eqnarray}
which is 1 in our convention. This equation together with the
condition (\ref{Dcon}) provides the correct tension of a D-string,
\begin{eqnarray}
 4\pi c_{\rm m} T_{\rm D3}= \frac{n}{g_{\rm s}}= n T_{\rm D1}.
\end{eqnarray}
Here $n$ should be a positive integer since the left hand side is
positive. We have shown that the remnant, represented by the
magnetic S3-brane solution, has the tension of a D-string, which
supports our claim that the remnant is a D-string.  A boundary
state discussion of this claim is also presented in Section
\ref{S3-D1bdstateSec}.

It is interesting to relate the above dualization procedure to a
discussion of S-duality. In fact Ref.~\cite{Tseytlin-Sduality}
discussed S-duality for D3-branes and used a Euclideanized
version of the D3-brane action for simplicity, which from our
viewpoint is an S3-brane action. As compared to our dualization
procedure Eq.~(\ref{multerm}), in the dualization process of
Ref.~\cite{Tseytlin-Sduality} the Lagrange multiplier field
$\phi_B$ enforcing the Gauss condition came with a factor of
``$i$''. The factor of ``$i$'' was argued to arise from the
Euclidean nature of the brane worldvolume.  The effect of this
alternate dualization procedure with an explicit factor of ``$i$'',
is that we reproduce the action in Eq.~(\ref{imaginaryFaction}).
Therefore this alternate dualization procedure is equivalent to
field strengths coming from the M-theory circle and not from the
non-BPS brane.

Finally, for this case the duality group discussed in
Section \ref{sis} becomes $SO(1,2)$ acting on $(X^0, \phi_B,
A_\chi)$, due to the ``$i$'' factor. The
 electric-magnetic duality is now
the more standard $SO(2)$ duality rotation, which is consistent
with the interpretation that this S3-brane decays into a D-string.
Interestingly, for the solution with the factor of ``$i$'', we can
ignore the $\chi$ direction and regard the solution as an
S2-brane instead of the S3-brane. This solution represents the
formation of a D0-brane from the S2-brane in type IIA theory. The
magnetic field was originally the scalar field for the M-theory
circle, thus this solution in the M-theory side represents a
lightlike particle emission process from the spacelike M2-brane.


\section{Strings and D-branes as Boosted S-branes}
\label{bsas}
\setcounter{footnote}{0}

A succinct summary of our characterization of S-branes so far is
that they are ways to follow the time dependent defect formation
process.
In this section we further discuss the $(p,q)$ strings of the
previous sections. We will find how certain ``boosted'' S1-branes
apparently become ordinary D-branes and fundamental strings
moving in the bulk. In fact, these boosted S1-branes extract late
time information of the remnant formation solutions which we
studied before. We start by presenting solutions of S1-brane
actions and discussing their properties.  The corresponding
tachyon solutions are then presented, and it is shown how in a
certain limit these solutions apparently become $(p,1)$ strings.
Boundary states for the boosted S-branes are also constructed, and
we show that they become boundary states of
$(p,1)$ strings in the limit relevant to the S-branes
discussed in the previous sections. Finally, we examine the
boundary state of the magnetic S3-brane in Section \ref{msfm}, and
show that at late times this solution produces a D-string
boundary state consistently.

\subsection{Boosted S1-branes}
\label{sitsa}

The S3-branes of Sections \ref{sfs} and \ref{dfs} eventually
confined into $1+1$ dimensional remnants so
it should be interesting to analyze S1-branes directly. Since the
``static'' S1-branes are spacelike in the target space we will
have to ``boost'' them to become timelike in the target space.
These boosted S1-branes are expected to be almost the same as the
spike solutions in Section \ref{sfs} and \ref{dfs} at late times,
except that the boosted S1-branes have at least one D1-brane
charge.

The general S1-brane action is
\begin{eqnarray}
  S =  \int
\! d^2x\; \sqrt{\det (\delta_{ij}-\p_i X^0 \p_j X^0 +
     F_{ij})}
\end{eqnarray}
where the Euclidean worldvolume is parametrized by $x^i$ with
$i=1,\chi$. First let us consider a solution relevant for the
fundamental string formation in Section \ref{sfs}.
As in the previous solutions, we turn on only $A_\chi$ among the gauge
fields and
assume $\p_\chi=0$, so the action simplifies to
\begin{eqnarray}
S =  \int
\!d^2x\; \sqrt{1-(\p_1 X^0)^2 +
     (\p_1 A_\chi)^2} \ .
\label{simplifies}
\end{eqnarray}
When the BPS-like relation $X^0=\pm  A_\chi$ holds, the equations
of motion become linear:
\begin{eqnarray}
  \p_1 \p_1 X^0=0.
\end{eqnarray}
This holds for any S$p$-brane if the above ansatz is applied,
and the spike solution of Section \ref{sfs} and Ref.~\cite{Sbraneaction}
was of this type.
In the present case $p=1$,
the solutions are simple
\begin{eqnarray}
  X^0=cx^1, \quad F_{1\chi}= c
\label{simps}
\end{eqnarray}
where the parameter $c$ describes the velocity in the target space
\begin{eqnarray}
  \frac{\p x^1}{\p X^0} = 1/c.
\end{eqnarray}
Due to the presence of the field strength $F_{1\chi}$
on the S-brane, the
resultant configuration can be timelike, $c>1$. The configuration
is a one dimensional object moving in the target space with speed
$1/c$ along the $x^1$ direction. If $c>1$, this object moves
slower than the speed of light and apparently becomes a physically
meaningful moving 1-brane!
The induced electric field on the
1-brane is
\begin{eqnarray}
  F_{0\chi} = \frac{\partial x^1}{\partial X^0 } F_{1\chi} =
1
\end{eqnarray}
which is the critical value.  If one tries to use a usual DBI
analysis for this moving 1-brane by assuming that this 1-brane is
a D1-brane, the DBI action becomes imaginary.  So although this
seems to be similar to a normal bound state of strings and
branes, this configuration seems to only have an S-brane
description using the S-brane action.

We can generalize this solution so that it deviates from the
BPS-like relation. A simple calculation shows that a generalized
solution is
\begin{eqnarray}
  \partial_1 X^0 = \frac{c_1}{\sqrt{1-c_2^2+c_1^2}},\quad
 F_{1\chi} = \frac{c_2}{\sqrt{1-c_2^2+c_1^2}}.
  \label{nbpspq}
\end{eqnarray}
In this case the induced electric field takes on arbitrary values
\begin{eqnarray}
  F_{0\chi} = \frac{c_2}{c_1},
\label{elec}
\end{eqnarray}
although we still have the restriction on the parameters $c_1$ and
$c_2$
\begin{eqnarray}
  1-c_2^2+c_1^2 \geq 0
\end{eqnarray}
coming from the reality condition for the S1-brane action.
The
velocity of the moving D1-brane has a lower bound related to the
field strength $F_{1\chi}$.
Expressing $c_2$ in terms of $F_{1\chi}$ and
$c_1$ as
\begin{eqnarray}
  c_2
=\frac{F_{1\chi}}{\sqrt{1 + (F_{1\chi})^2}}
\sqrt{1 + c_1^2},
\end{eqnarray}
it is not difficult to see that
\begin{eqnarray}
  \Biggm|\frac{\p x^1}{\p x^0}\Biggm|
= \frac{\sqrt{1+c_1^2}}{c_1} \frac{1}{\sqrt{1+(F_{1\chi})^2}}
\geq \frac{1}{\sqrt{1+(F_{1\chi})^2}}.
\label{lb}
\end{eqnarray}
Setting $c_1=c_2=c$ brings us back to the BPS solution (\ref{simps}).

These solutions include ones which describe static configurations
in the bulk.  Setting the velocity to zero in
Eq.~(\ref{nbpspq}), we get the relationship $c_2^2 = 1+c_1^2$ and in
this case the induced electric field can be larger than the
critical value
\begin{eqnarray}
  F_{0\chi} = \frac{\sqrt{1+c_1^2}}{c_1}
\geq 1 \ .
\label{larger}
\end{eqnarray}
Again, we see that this static one-dimensional object exceeds the
validity of the usual DBI action, unless $c_1=\infty$. In the
limit $c_1=\infty$ the configuration is static and has a critical
electric field so this configuration can also be described by the
usual D1-brane action. However this limit is rather singular and
it apparently represents an (n,1) string with
$n\rightarrow\infty$. We identify this as an infinite number of
fundamental strings where the D1-brane effect has disappeared
\cite{Polyakov}. On the other hand, the limit $c_1\sim c_2 =
\infty$ is just like the late time behavior of the spike solution
found in Section \ref{sfs} and Ref.~\cite{Sbraneaction} so here we
see a nice agreement between these two S-brane solutions.

\subsection{Tachyon condensation representation}

\label{tcr}

Our general S-brane analysis is based on the belief that any
solution of the S-brane action has a corresponding tachyon
solution on an unstable brane.  The solution given in the
previous subsection should hence have a tachyon description.
Since the solution is just a boosted S-brane, it is natural to
expect that the corresponding tachyon solution can be generated
by the worldvolume boost from the homogeneous rolling tachyon
solution. In this case, one has to perform a Lorentz boost
respecting the open string metric. Let us see this in more detail.

We start with the following general Lagrangian for a non-BPS
D2-brane,
\begin{eqnarray}
  L = -V(T)\sqrt{-\det (\eta + F)}\;{\cal F}\left(
G^{\mu\nu} \partial_\mu T\p_\nu T \right),
\label{tacacg}
\end{eqnarray}
where ${\cal F}$ is a function defining the kinetic energy
structure of the tachyon, and $G^{\mu\nu}$ is the open string
metric. This action is the general form for the linear tachyon
profiles.
Almost all the Lagrangians which have been
investigated so far, such as Sen's rolling Lagrangian
\cite{roll,tact},
BSFT \cite{BSFT, BSFTsoliton,BSFTRR}, and
Minahan-Zwiebach models \cite{Zwie},
are included in this general form. Let
us examine the tachyon field which depends only on $x^0$ and
$x^1$. If one chooses a gauge $A_\chi=0$ and turns on only $A_1$,
then the gauge field equations of motion are satisfied trivially
for the constant gauge field strength $F_{1\chi}$. Then the
problem reduces to the situation where we have to solve only the
tachyon equation of motion under the background of the field
strength which appears only in the open string metric. In our
case the explicit form of the inverse open string metric is
\begin{eqnarray}
  G^{\mu\nu} = {\rm diag}
\left( -1,\frac{1}{1+(F_{1\chi})^2},
\frac{1}{1+(F_{1\chi})^2} \right).
\end{eqnarray}
where $\mu=0,1,\chi$.  The metric in the
$x^0$-$x^1$ spacetime is
\begin{eqnarray}
  G_{\mu\nu} = {\rm diag}(-1, 1+(F_{1\chi})^2).
\end{eqnarray}
The simplest solution is a homogeneous solution, $\p_1 T=\p_\chi
T=0$.  Since in this case we turned on only the magnetic field, we
have that $G_{00}=-1$, and so this solution is just the same as
the one with vanishing field strength. One can integrate the
equations of motion for $T$ and then obtain a
solution\footnote{At this stage we exceed the validity of the
BSFT
tachyon action (\ref{tacacg}) since the solution is not linear in
$x^0$ \cite{BSFTmatter}.} $T=T_{\rm cl}(x^0)$. Without loss of
generality, we may assume that the tachyon passes the top of its
potential at $x^0=0$, i.e. the equation $T_{\rm cl}(x^0)=0$ is
solved by $x^0=0$.

\begin{figure}[ht]
\begin{center}
\begin{minipage}{12cm}
\begin{center}
\includegraphics[width=10cm]{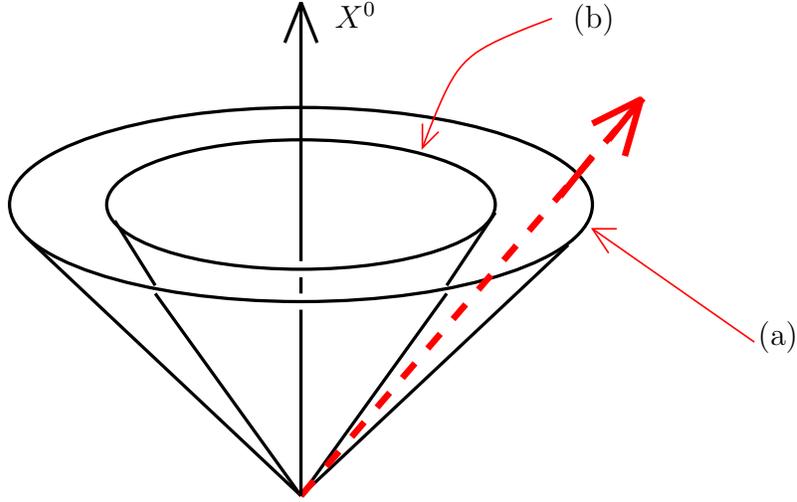}
\put(0,60){(a)} \put(-160,180){$X^0$} \put(-70,180){(b)}
\end{center}
\caption{The light cone structure on the non-BPS D2-brane
worldvolume. The closed string light cone (a) is always located
outside the open string
 light cone (b). The dashed line denotes the motion of the boosted
 S-brane which is both timelike with respect to the closed string light
 cone and spacelike with respect to the open string light cone. }
\label{lightcones}
 \end{minipage}
\end{center}
\end{figure}

We next perform a Lorentz boost in the 01 spacetime directions
which preserves the open string metric. For this purpose we
define a rescaled coordinate $\tilde{x}^1\equiv
\sqrt{G_{11}}x^1$.  In these rescaled coordinates the metric
becomes $\tilde{G}_{\mu\nu}={\rm diag}(-1,1)$ and the Lorentz
boost takes the usual form
\begin{eqnarray}
  \left(
    \begin{array}{c}
x^0 \\ \tilde{x}^1
    \end{array}
\right) \rightarrow
  \left(
    \begin{array}{c}
{x^0}' \\ {\tilde{x}^1}{}'
    \end{array}
\right) =
  \left(
    \begin{array}{cc}
\cosh\gamma & \sinh\gamma \\
\sinh\gamma &\cosh \gamma
    \end{array}
\right)
  \left(
    \begin{array}{cc}
x^0 \\ \tilde{x}^1
    \end{array}
\right).
\end{eqnarray}
The line where the original defect is located, $x^0=0$, is
boosted to a tilted line
\begin{eqnarray}
  x^0 + \tanh\gamma \sqrt{G_{11}} x^1=0
\end{eqnarray}
so the defect is now moving along the $x^1$ direction with
velocity
\begin{eqnarray}
  \frac{\p x^1}{\p x^0} = \frac{-1}{\sqrt{G_{11}}\tanh\gamma}.
\end{eqnarray}
The important point here is that the absolute value of this
velocity can be made less than unity. By definition
$|\tanh\gamma|\leq 1$, so if the field strength vanishes, the
velocity of the configuration is greater than that of light; the
worldvolume of the defect is still spacelike.  If we turn on a
constant field strength, then a large boost will
make the defect timelike. This property is a direct result of the
fact that the open string light cone lies inside the closed
string light cone \cite{gib}. Due to this fact one may obtain
timelike D-branes from Spacelike-branes (see
Fig.~\ref{lightcones}).

The lower bound for the velocity of the moving D-brane (\ref{lb})
should be seen also in this tachyon solution. In fact, it is
given by
\begin{eqnarray}
\biggm|\frac{\p x^1}{\p x^0}\biggm| \; \geq\;
\Bigm|1/\sqrt{G_{11}}\Bigm|\;  =
\;\frac{1}{\sqrt{1+(F_{1\chi})^2}},
\end{eqnarray}
which coincides with (\ref{lb}).  For the S-brane, the limit
$F_{1\chi}\rightarrow \infty$ makes the S-brane worldvolume
static. Let us study what happens to the tachyon solution in this
limit. The boosted tachyon configuration is
\begin{eqnarray}
 T= T_{\rm cl}\left((\cosh \gamma)
x^0+(\sinh \gamma) \sqrt{G_{11}}x^1\right).
\label{boostsol}
\end{eqnarray}
The original solution $T_{\rm cl}(x^0)$ has the rolling tachyon
behavior for large $x^0$,
$T_{\rm cl}\sim  x^0$.   Therefore in the limit $F_{1\chi}\rightarrow
\infty$, this boosted tachyon solution becomes
\begin{eqnarray}
 T \sim u x^1, \quad u \equiv  F_{1\chi} \sinh \gamma\rightarrow
\infty
\label{boostedbecome}
\end{eqnarray}
and this linear dependence on $x^1$ coincides with the familiar
static D-string kink solution. The coefficient of the linear term
diverges, which is also consistent with the BSFT renormalization
argument for D-brane kink solutions \cite{BSFTsoliton, BSFTRR}.

It is clear that the moving 1-brane has unit D-string
charge. Taking into account that the integration surface
enclosing the defect in the original non-BPS brane worldvolume
is not necessarily timelike, the S-brane
charge is just the same as the D-brane charge \cite{stro}.
So, if the S-brane
worldvolume is deformed to be timelike, it should give
ordinary D-brane charge. This can be easily seen from the
RR-tachyon coupling in the non-BPS brane \cite{BSFTRR},
\begin{eqnarray}
 \int C \wedge dT e^{-T^2}.
\end{eqnarray}
Here $dT$ can be evaluated as
\begin{eqnarray}
 dT = \frac{\p T_{\rm cl}({x^0}')}{\p {x^0}'} d({x^0}').
\end{eqnarray}
Therefore if the boosted line ${x^0}'=0$ becomes timelike, usual
D-brane charge is generated in which the RR source is distributed
on a hypersurface timelike in the bulk closed string metric.

Here we stress that the boosted tachyon configuration has usual
D-string charge, so the configuration should represent an $(n,1)$
string with $n\rightarrow \infty$, as seen in Section
\ref{sitsa}. Then, how is the fundamental string charge $n$ seen in the
tachyon description? The answer is that the fundamental string
charge is expected to be realized only in the induced electric
field, not in the tachyon field. In fact, if we recall the
noncommutative soliton representing a fundamental string
\cite{NCfun}, there the tachyon sits at the bottom of the
potential from the first place. In the present case using
Eq.~(\ref{boostsol}), it is easy to evaluate the induced electric
field
\begin{eqnarray}
 F_{0\chi}= \frac{\p x^1}{\p x^0}F_{1\chi} =
-\frac{F_{1\chi}}{\sqrt{G_{11}} \tanh\gamma}
\stackrel{\mbox{\scriptsize $F_{1\chi}\rightarrow \infty$}}
{\longmapsto} -\coth\gamma
 \label{inducedE}
\end{eqnarray}
and we find that this agrees with the S1-brane analysis
Eq.~(\ref{larger}).  So in the limit $\gamma \rightarrow \infty$
we have a critical electric field $F_{0\chi}=-1$.

In addition to the charges of the defects, their energy is
another important physical quantity to study. Though one expects
that the energy of the boosted S-brane should depend on the
tension of the S-branes whose precise value is unknown, we may
proceed by using the explicit expression of the corresponding
tachyon solution. The detailed analysis is presented in
Appendix \ref{appb}.

Although we have just seen how the S1-brane seems to describe
$(n,1)$ string bound states, one might question the validity of
the solutions since the S-brane solutions allow for faster than
light travel.  Let us examine the tachyon configurations to see
how this occurs. As discussed around Eq.~(\ref{boostedbecome}) a
static brane has zero width while all moving configurations
acquire a finite width. When the width is small relative to the
background it is easy to say that there is lump which is actually
moving, and in such cases the lump is moving slower than the
speed of light. If we speed up the configuration, its width
increases and the lump in the tachyon field becomes hard to
separate from the background. In such cases it is difficult to
say if the lump is moving and instead we should describe the
configuration as a collective motion of the tachyon field which
just resembles a lump moving. When the S-branes move faster than
light, the configurations do not have good interpretations in
terms of lumps or branes in motion and so it is okay if the
configuration ``moves'' at a speed greater than light.

\subsection{Boundary state and fundamental string charge}

In the previous section it was shown that the boosted S-brane
carries D-string charge and the tachyon configuration had the
usual D-string form. However, since an electric field is induced
on this D-string as shown in Eq.~(\ref{elec}), the 1-brane is
expected to be an $(n,1)$ string which also possesses fundamental
string charge. The easiest way to see if this object carries such
a charge is to study its boundary state, especially its coupling
to the bulk NS-NS gauge field. In this subsection we explicitly
construct a boundary state for the boosted S-branes of
Eq.~(\ref{nbpspq}).

According to Gutperle and Strominger \cite{stro}, the boundary
state for an S$p$-brane\footnote{In the following, we identify
our flat S-brane in the rolling tachyon context with the SD-brane
which is defined to be a brane on which open strings can end with
Dirichlet boundary conditions along time. } satisfies the
following boundary conditions
\begin{eqnarray}
  (\alpha_n^\mu + {\cal O}^{\mu}_{\;\nu}\tilde{\alpha}^\nu_{-n})
|B,\eta\rangle =0
\label{bsd}
\end{eqnarray}
(and similar expressions for the worldsheet fermions).  The
orthogonal matrix ${\cal O}$ is given by
\begin{eqnarray}
{\cal O}^\mu_{\;\nu}= {\rm diag} (-1,1,\cdots,1,-1,\cdots,-1)
\label{s1b}
\end{eqnarray}
where we have $p+1$ entries giving $+1$, specifying the Neumann
directions.   For
spacelike branes the first entry ${\cal O}^0_{\;0}$ is negative
due to the Dirichlet boundary condition for the time direction.

We now proceed to find the boundary state for the boosted
S1-brane. Since our solution has constant field strength and
constant velocity, it is expected that only the orthogonal matrix
${\cal O}$ will be modified.\footnote{Also the normalization of
the boundary state, which is usually identified with the DBI
Lagrangian, will be modified but in this paper we will not
consider this point.} We work out the bosonic string case for
simplicity. The worldsheet boundary coupling in the string sigma
model should be
\begin{eqnarray}
\oint d\tau \left( F_{1\chi} X^1 \frac{\partial}{\p \tau} X^\chi
+ V X^1 \frac{\p}{\p\sigma}X^0 \right)
\end{eqnarray}
where $V$ is the inverse of the velocity of the moving D1-brane,
while we normalize the bulk action as
\begin{eqnarray}
  \frac12 \int\! d\sigma d \tau \;
\p_a X^\mu \p_b X^\nu \eta^{ab}\eta_{\mu\nu}
\end{eqnarray}
with the oscillator expansion
\begin{eqnarray}
  X^\mu = x^\mu + p^\mu \tau \frac{i}{2}+
\sum_{n\neq 0} \frac1{n} \left( \alpha_ne^{in(\sigma + \tau)}
+\tilde{\alpha}_ne^{in(\sigma - \tau)} \right).
\end{eqnarray}
The variation of the action gives the boundary conditions at
$\sigma=0,\pi$ as
\begin{eqnarray}
&&  \p_\sigma X^1 - F_{1\chi} \p_\tau X^\chi - V\p_\sigma X^0=0,
\nonumber\\
&& \p_\sigma X^\chi +F_{1\chi} \p_\tau X^1 =0,
\label{bccond}\\
&& \p_\tau X^0 - V \p_\tau X^1=0.
\nonumber
\end{eqnarray}
The last condition is due to the original Dirichlet boundary
condition for the time direction $X^0$. Substituting
\begin{eqnarray}
  \p_\sigma X^\mu \biggm|_{\sigma=0} =
-\frac12 \sum_n (\alpha_n^\mu+  \tilde{\alpha}_{-n}^\mu),
\quad
  \p_\tau X^\mu \biggm|_{\sigma=0} = -\frac12 \sum_n
(\alpha_n^\mu-  \tilde{\alpha}_{-n}^\mu)
\end{eqnarray}
into the above boundary conditions (\ref{bccond}), we obtain
\begin{eqnarray}
&&  \alpha_n^0 - \tilde{\alpha}_{-n}^0
- V (  \alpha_n^1 - \tilde{\alpha}_{-n}^1)=0,
\nonumber\\
&&  \alpha_n^1+ \tilde{\alpha}_{-n}^1 - F_{1\chi} (
\alpha_n^\chi - \tilde{\alpha}_{-n}^\chi)
- V (  \alpha_n^0 + \tilde{\alpha}_{-n}^0)=0,
\nonumber\\
&&  \alpha_n^\chi+ \tilde{\alpha}_{-n}^\chi + F_{1\chi} (
\alpha_n^1 - \tilde{\alpha}_{-n}^1)=0.
\end{eqnarray}
Solving these equations, we obtain a new orthogonal
matrix specifying the
boundary condition
\begin{eqnarray}
  \tilde{\cal O}^\mu_{\;\nu} =
\frac{1}{1+F_{1\chi}^2-V^2}
\left(
  \begin{array}{ccc}
-(1+F_{1\chi}^2+V^2) & 2V & 2F_{1\chi}V \\
-2V & 1-F_{1\chi}^2+V^2 &2F_{1\chi} \\
2F_{1\chi}V& -2F_{1\chi} & 1-F_{1\chi}^2-V^2
  \end{array}
\right),
\label{ot}
\end{eqnarray}
where $\mu,\nu=0,1,\chi$.
It should be noted here that off-diagonal entries appear in
$\tilde{\cal O}$, and these are responsible for the fundamental
string charge.  There is now a non-vanishing overlap of the
boundary state with a NS-NS $B$-field state $|B_{\mu\nu}^{\rm
NSNS}\rangle$. This represents a source for the $B$-field
\begin{eqnarray}
  \langle B| B^{\rm NSNS}_{0\chi}\rangle \propto
 -\tilde{\cal O}_{0\chi}+\tilde{\cal O}_{\chi 0}
=\tilde{\cal O}^0_{\;\chi}+\tilde{\cal O}^\chi_{\;0}\neq 0.
\end{eqnarray}
Here we lowered the indices by $\eta_{\mu\nu}$ which appears in
the oscillator commutation relations. This shows that the moving
D-string carries fundamental string charge and becomes a source
for the target space NSNS $B$-field.

To gain a better understanding of this source, such as the amount
of charge $n$ it has, let us study the structure of the orthogonal
matrix ${\cal O}$ in more detail. We started from a S1-brane
boundary state (\ref{s1b}) which has a Dirichlet boundary
condition along time and then boosted it to obtain the matrix in
Eq.~(\ref{ot}). This can be compared with the ordinary $(n,1)$
string boundary state constructed in Ref.~\cite{CallanK} which is
obtained from the boundary state of a {\it D1-brane} by
introducing the boundary coupling\footnote{ Here we changed the
notation from Ref.~\cite{CallanK} as $\sigma \leftrightarrow
\tau$ and $(0,1,2) \rightarrow (0,\chi,1)$ to fit our
computation, and used $-v$ instead of $V$ in Ref.~\cite{CallanK}
to avoid confusion.}
\begin{eqnarray}
 \oint \! d\tau \left(
EX^0 \frac{\p}{\p \tau}X^\chi - v X^0 \frac{\p}{\p\sigma}X^1
\right).
\end{eqnarray}
The orthogonal matrix obtained in Ref.~\cite{CallanK} was
\begin{eqnarray}
  \tilde{\cal O}^\mu_{\;\nu} =
\frac{1}{1-E^2-v^2}
\left(
  \begin{array}{ccc}
1+E^2+ v^2 & -2v & 2E \\
2v & -1+E^2-v^2 & 2vE \\
2E& -2vE & 1+E^2-v^2
  \end{array}
\right)
 \label{CK}
\end{eqnarray}
and the associated boundary state describes an $(n,1)$ string
moving with the speed $v$ along the $x^1$ direction.
 The charge $n$ is given by the electric flux
on the worldvolume theory,
\begin{eqnarray}
 n = \frac{E}{\sqrt{1-E^2-v^2}}.
\end{eqnarray}

Remarkably, the matrix
(\ref{ot}) is identical with (\ref{CK})
under the relation
\begin{eqnarray}
V = \frac{1}{v}, \quad  F_{1\chi} = \frac{E}{v}.
\label{relf}
\end{eqnarray}
This is indeed what we expected since the first equation is just
$v=\p x^1/\p x^0= 1/V$ and the second equation is just the change
of the coordinates for $E=F_{0\chi}$ which we have found in
previous subsections. This suggests that the boosted
S-brane boundary state (\ref{ot}) describes a moving $(n,1)$
string, but in a strict sense this is not the case. Let us compare
the regions of parameter space where the actions are valid. The
description (\ref{ot}) is valid if the S-brane Lagrangian is real,
\begin{eqnarray}
 1+F_{1\chi}^2-V^2 \geq 0.
\end{eqnarray}
Substituting the identification (\ref{relf}) into the above inequality,
we find
\begin{eqnarray}
 1-E^2-v^2\leq 0,
\end{eqnarray}
which is the region where the description (\ref{CK}) is invalid
since the D1-brane Lagrangian becomes imaginary. Therefore,
although the boundary states have the same structure, their valid
regions of parameter space are different. The two descriptions
overlap only in the case of vanishing Lagrangians where the
fundamental string charge $n$ goes to infinity. This means that
the fundamental string (limit) can be described by both the boosted
S1-brane and the D1-brane!

In the static case we can see this correspondence more directly.
In the S-brane boundary conditions (\ref{bccond}), we take the
limit
\begin{eqnarray}
 E = \frac{F_{1\chi}}{V}\rightarrow 1,
\quad v = \frac{1}{V}\rightarrow 0
\end{eqnarray}
which is expected to give static fundamental strings. Then
Eq.~(\ref{bccond}) reduces to
\begin{eqnarray}
\p_\tau X^1=0, \quad
\p_\tau X^\chi + \p_\sigma X^0=0.
\end{eqnarray}
The first equation tells us that the object has Dirichlet boundary
condition along $x^1$ and so it has worldvolume along $x^0$ and
$x^\chi$, while the second equation is the $|E|=1$ limit of the
mixed boundary condition on a D-string,
\begin{eqnarray}
 F_{0\chi}\p_\tau X^\chi + \p_\sigma X^0=0.
\end{eqnarray}
So this is precisely the fundamental string limit.

\subsection{S-brane description and T-duality}

At this stage it is very natural to ask, ``What is the boosted
S-brane without taking the fundamental string limit ($=$
vanishing Lagrangian limit)?'' To approach a possible answer to
this question, let us observe what happens to the orthogonal
matrix in the boundary state.  For simplicity we examine the
static case. The boundary state of a static $(n,1)$ string
presented in Ref.~\cite{CallanK} is defined through its
orthogonal matrix
\begin{eqnarray}
  \tilde{\cal O}^\mu_{\;\nu}(E) =
\frac{1}{1-E^2}
\left(
  \begin{array}{cc}
1+E^2 & 2E   \\
2E & 1+E^2
  \end{array}
\right),
\label{CK2}
\end{eqnarray}
where $\mu,\nu=0,\chi$. Here of course $E$ should be less than or
equal to 1. On the other hand, the boosted S1-brane with the
static limit $V=\infty$ is also described by the above matrix
with $E\geq 1$. To relate these two descriptions, we see that if
perform the transformation
\begin{eqnarray}
 E \rightarrow \tilde{E} = 1/E,
\end{eqnarray}
then the matrix $\tilde{\cal O}$ transforms as
\begin{eqnarray}
 \tilde{\cal O}(\tilde{E}) = -\tilde{\cal O}(E).
\end{eqnarray}
Interestingly, this means that the case with electric field $E$
larger than 1 is related to an $E$ smaller than 1 only by a sign
change of $\tilde{\cal O}$. The change in the sign of $\tilde{\cal
O}$ is equivalent to the replacement $\tilde{\alpha} \rightarrow
-\tilde{\alpha}$ which is a T-duality along $x^0$ and $\chi$
directions, see Eq.~(\ref{bsd}).

So what we have found here is that the description of $E$ larger
than 1 can be obtained by T-duality along $x^0$ and $\chi$. Let
us discuss the meaning of this duality more. Before examining our
present case, it is instructive to remember the ordinary
T-duality along spatial directions for D-branes.
Let us consider a bound state of $n$ D0-branes and $m$
D2-branes. The D2-brane worldvolume is extended along $x^1$ and
$x^2$. The density of the D0-branes per unit area on the
worldvolume of a single D2-brane is just the magnetic field
induced on the D2-brane, $F_{12}=n/m$. The open string boundary
condition becomes a mixed boundary condition. Now let us take a
T-duality along $x^1$ and $x^2$. First, T-dualizing along $x^1$
transforms this D2-D0 bound state to a D1-brane winding the 12
torus $n$ times along $x^1$ and $m$ times along $x^2$. Second,
take the T-duality along $x^2$. We then get a bound state of $n$
D2-branes and $m$ D0-branes, giving an induced magnetic field
$\tilde{F}_{12}=m/n=(F_{12})^{-1}$. This shows that the inversion
of the magnetic field can be understood as T-duality.

Let us apply this well-known idea to our case, and see what
happens to a $(n,1)$ string when we T-dualize along
$x^0$ and $\chi$. Consider a static $(n,1)$ string
stretched along the $\chi$ direction. The induced electric field
$E=F_{0\chi}<1 $ parametrizes the number of bound fundamental
strings.  First let us take a T-duality along $\chi$. The
resultant configuration is a D0-brane moving at the speed $E$ which does
not exceed the speed of
light. This moving D0-brane can be thought of as a ``winding''
D0-brane, that is, a D0-brane winding $1/E$ times along $x^0$ and
$1$ time along $\chi$. The winding along $\chi$ should be thought of as
an S0-brane since the worldvolume is only along this spatial
direction. Now
take a second T-duality along $x^0$. The former $1/E$
D0-brane becomes $1/E$ S(-1)-branes, while the latter S0-brane
becomes a single D1-brane. Therefore, after the T-dualities,
we have a bound state of a single D1-brane and $1/E$
S(-1)-branes. This statement is very plausible in view of how we
derived the boosted S-brane: there we considered an S1-brane with
magnetic field $F_{1\chi}$, that is exactly a bound state of an
S1-brane and S(-1)-branes.  If we consider now the boosted S-brane so the
S1-brane is timelike, i.e. a D1-brane, the resultant
object should be a bound state of a D1-brane and S(-1)-branes.

Since the S-brane description in the previous subsections is valid
for $E\geq 1$, the case $E=1$ is the only overlapping region and
has two equivalent descriptions. However, the above observation
leads us to an intriguing conjecture:  Any $(n,1)$ string can be
thought of as a bound state of a D1-brane and $E$ S(-1)-branes
with $E<1$.  Here we do not specify how the latter bound state
should be described but there might be some advantages in
treating the $(n,1)$ strings from the S-brane point of view. To
illustrate this point, consider the RR coupling on a D$p$-brane
\begin{eqnarray}
 \int C^{(p+1)} + F \wedge C^{(p-1)} + \cdots.
\end{eqnarray}
Let us turn on a constant
electric field $E_{01}$. Usually this is said to turn the D$p$-brane
into an (F, D$p$) bound state, but what does
the above RR coupling tell us? The second term gives
\begin{eqnarray}
 E_{01} \int C^{(p-1)}_{23\cdots p}.
\end{eqnarray}
This is a source term for the RR $(p\!-\!1)$-form with spatial
indices, or in other words for an S$(p\!-\!2)$-brane. This
suggests that the fundamental strings can be thought of as smeared
S-branes, at least in the worldvolume of other mother D-branes in
which the fundamental strings are bound.

\subsection{Relation between S- and D-brane descriptions}

In the above we have learned that while D-branes with small
electric fields are described by D-brane actions, D-branes with
large electric fields are described by S-brane actions. Following
the previous subsection, here we further explore the T-duality
which interchanges these two classes of configurations.

For simplicity, only the electric field
in the $\chi$ direction is turned on.
The Lagrangian, electric flux density
and the Hamiltonian for the D-brane are given by
\begin{equation}
L = -\sqrt{1-E^2}, \quad
D = \frac{E}{\sqrt{1-E^2}}, \quad
H = \frac{1}{\sqrt{1-E^2}};
\end{equation}
and those for the S-brane are
\begin{equation}
L = \sqrt{-1+E^2}, \quad
D = \frac{E}{\sqrt{-1+E^2}}, \quad
H = \frac{1}{\sqrt{-1+E^2}}.
\end{equation}
The range of electric fields valid for the D-brane description is
$E^2 < 1$, which is mapped to the range of validity $E^2 > 1$ for
the S-brane description by the T-duality along the time direction
\begin{equation}
E \rightarrow \frac{1}{E}
\end{equation}
considered in the previous subsection. From
the expressions above,
we find that this map induces
the interchange of $D$ and $H$,
or equivalently the interchange of
the fundamental string charge and the energy.
Recall that ordinary T-duality interchanges
winding modes with Kaluza-Klein modes.
Since the total string number can be thought of
as the ``winding number'',
and the energy as the ``momentum'' in the time direction,
roughly speaking, the interchange of $D$ and $H$
is what one would expect for the T-duality in the time direction.

\subsection{Boosted S3-brane as a D-string}
\label{S3-D1bdstateSec}

Earlier in this section we saw how the late time part of the
solution of Section \ref{sfs} can be realized as a boosted
S1-brane. We may expect that in the same manner the late time
configuration of the spike solution of Section \ref{msfm} can also
be obtained as a boosted S-brane. Here we will present a boosted
solution of an S3-brane action with magnetic fields,\footnote{Though so
far in this section we have used S1-branes, in this subsection
we need magnetic fields and so use an S3-brane instead.} and show
that actually the boundary state of the boosted S3-brane reduces
to that of a static D-string.

As explained in Section \ref{msfm} we may consider field strengths
on the S3-brane arising from the excitations of a scalar field
along the M-theory circle. If we assume that the fields in
(\ref{spvortex}) are independent of $x^2$, $x^3$ as well as
$x^4$, we obtain for vanishing $A_4$ ($=X^4$)
\begin{eqnarray}
 L = \sqrt{1-(\p_1 X^0)^2 + (\p_1 X^{10})^2} \ .
\label{laggen2}
\end{eqnarray}
The field $X^{10}$ is related to the original field strength $B_1
\equiv \tilde{F}_{23}$ through the Legendre transformation,
\begin{eqnarray}
 \frac{\delta}{\delta B_1}
\left[ \sqrt{1-(\p_1 X^0)^2 - B_1^2 + (B_1 \p_1X^0)^2} -
B_1\p_1X^{10} \right]=0
\end{eqnarray}
where the factor of $i$ has been included as discussed earlier.
This is rewritten as
\begin{eqnarray}
 \p_1 X^{10} = -B_1 \sqrt{\frac{1-(\p_1 X^0)^2}{1 - B_1^2}}
\end{eqnarray}
so that the S3-brane can become a timelike
object, $|\p_1 X^0|>1$.

The Lagrangian (\ref{laggen2}) has the same form as
(\ref{simplifies}), as it should due to S-duality.  There exists
a general solution similar to (\ref{nbpspq}),
\begin{eqnarray}
 \p_1 X^0 = \frac{c_1}{\sqrt{1-c_2^2+c_1^2}},
\quad
 \p_1 X^{10} = \frac{c_2}{\sqrt{1-c_2^2+c_1^2}}.
\end{eqnarray}
Let us take the BPS limit $c_1=c_2$ and furthermore the static
limit $c_1 \rightarrow \infty$. This is expected to be a D-string
since this limit provides the late time behavior of the spike
solution in Section \ref{msfm}. To check this, let us again look
at the worldsheet boundary condition of an attached fundamental
string. The appropriate inclusion of the boundary coupling leads
to\footnote{Although there appears ``$i$'' in this expression,
this might be absorbed into the redefinition of the worldsheet
variables.}
\begin{eqnarray}
&&  \p_\sigma X^2 - i \tilde{F}_{23} \p_\tau X^3 =0, \quad
 \p_\sigma X^3 +i \tilde{F}_{23} \p_\tau X^2 =0,
\\
&& \p_\tau X^0 - V \p_\tau X^1=0, \quad V \p_\sigma X^0 -
\p_\sigma X^1=0,
\end{eqnarray}
where $V$ is defined to be the value of $\p_1 X^0$ in the
solution as before. In the static limit, $V\rightarrow \infty$ and
$\tilde{F}_{23}\rightarrow \infty$, the above boundary conditions
reduce to
\begin{eqnarray}
\p_\tau X^3 = \p_\tau X^2 = \p_\tau X^1=0, \quad \p_\sigma X^0=0.
\end{eqnarray}
Remembering that we have a Neumann boundary condition for $x^4$,
this is precisely a boundary condition for a D-string extended
along $x^4$.

This analysis provides more evidence for the claim that the late
time remnant of the solution in Section \ref{msfm} is just a D-string.
Here we demonstrated that D-strings can be described by an
S3-brane, suggesting another interesting duality.


\section{S-brane and D-brane Interactions}
\label{sadi}

In this section we discuss how the formation of a codimension one
D-brane can be understood using an S-brane description of brane
creation.
In comparison, the solutions in Section~\ref{s3bwmf} describe the
formation of a $(p,q)$ string from an S3-brane which is defined
to be a spacelike defect on a non-BPS D4-brane.  On the non-BPS
D4-brane, these S-brane solutions are therefore describing the
formation of codimension three defects.  However, the simplest
case should be formation of a codimension one D-brane, which has
been studied in some literature \cite{Sentimeevole,
Mukohyamakinkform,LNT,Clinekinkform,IshidaUehara}.

Here we make a preliminary discussion of the interesting role which
S-branes play in RR charge conservation.  Our main point is that
in order to create charged defects we must also have charged S-branes
whose time dependent charge represents specific inflow and
outflow of charge into the system.  In a time evolution
transition, for example, we will discuss how RR charge can be
thought to be ``added'' by the S-brane
\begin{equation}
A \ \mbox{(with \ charge \ $q_1$)} \ \ \stackrel{
\mbox{\scriptsize S-brane``charge''}q_2}{\longmapsto}
B \ \mbox{(with \ charge \ $q_1+q_2$)}  \ .
\end{equation}

An interesting candidate process to examine is the time dependent
formation of a kink, see also Refs.~\cite{Sentimeevole,
Mukohyamakinkform,LNT,Clinekinkform,IshidaUehara}.  For simplicity
consider a kink D0-brane on a non-BPS D1-brane system. The kink
solutions for a D0-brane and the anti-kink solution for a
$\overline{\rm D0}$-brane are schematically
\begin{equation}
T_D(x)= \left\{ \begin{array}{l}
                                >0 \ \; {\rm for} \ x>0 \\
                                <0 \ \; {\rm for} \  x<0
             \end{array} \right. \hspace{.3in}
T_{\overline{D}}(x)= \left\{ \begin{array}{l}
                                <0 \ \; {\rm for} \ x>0 \\
                                >0 \ \; {\rm for} \  x<0 \ .
             \end{array} \right.
\end{equation}
Consider now a transition from kink to anti-kink.  This is a
configuration where the absolute values of the tachyon field
decrease and then increase again.  The crucial point is that there
should be a transition in the entire tachyon profile as it goes
through zero.  The time evolution of the configuration should
roughly pass through
\begin{equation}
T(x)= 0 \quad {}^{\forall} x
\end{equation}
which is flat!  Since the S-brane always appears in such a
transition, we attempt to ascribe the change in charge as being
due to the S-brane.  Although from the point of view of the
effective theory the S-brane is a very non-localized
instantaneous charged object, the complete tachyon profile paints
a more standard picture which shows that the transition is not
instantaneous. We will see however the consistency and simplicity
of the S-brane picture.

To go from kink to anti-kink, the S-brane must have charge two,
one to annihilate with the $\overline{\rm D0}$-brane
and one to create the
D0-brane.  The fact that a flat S-brane describes such a process
is very surprising as it is so simple and is different from our
other S-brane solutions.
Also as discussed in Ref.~\cite{Clinekinkform}, many branes and
anti-branes can be essentially created from a flat $T=0$ initial
condition.  It seems then that a flat charge one S-brane can
either destroy a D0-brane, or destroy a D0-brane and also create
equal numbers of branes and anti-branes.  If this statement were
true it would greatly reduce the usefulness of S-branes since
each S-brane would represent an infinite number of qualitatively
different processes. Fortunately, we shall see by considering
things more carefully that this is not the case and our
consideration here was too naive.  In fact we can consistently
conserve RR charge in the tachyon condensation process by
properly accounting for the S-branes.

\begin{figure}[bhtp]
\begin{center}
\includegraphics[width=10cm]{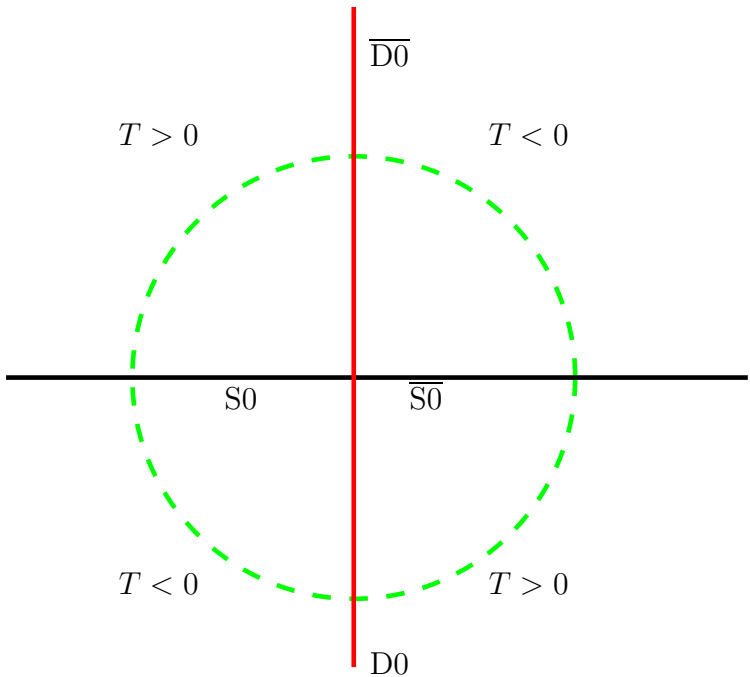}
\put(-200,100){S0} \put(-145,230){$\overline{\rm D0}$} \put(-145,0){D0}
\put(-130,100){$\overline{\rm S0}$}
\put(-100,30){$T>0$} \put(-100,200){$T<0$}
\put(-240,30){$T<0$} \put(-240,200){$T>0$} \caption{Formation of
an anti-kink using a kink and S-branes.} \label{charge1}
\end{center}
\end{figure}

Fig.~\ref{charge1} illustrates the time dependent kink
formation process and represents the entire non-BPS D1-brane
worldvolume with the vertical and horizontal directions
corresponding to time and space, respectively.  The horizontal
line $t=0$ indicates the location of the S0-brane, the upper half
vertical line is a $\overline{\rm D0}$-brane and the lower half
vertical
line is a D0-brane. For $t<0$, $T(x)>0$ for $x>0$ while $T(x)<0$
for $x<0$. For $t>0$, $T(x)<0$ for $x>0$ while $T(x)>0$ for $x<0$.

Although the horizontal line marks the $T=0$ region, it actually
consists of an S0-brane and an $\overline{\rm S0}$-brane.
The S0-brane is located at
$x<0$, $t=0$ while the $\overline{\rm S0}$-brane is at $x>0$,
$t=0$. This is clear
if we look at the tachyon configuration at $t=0$ since
$\dot{T}<0$ for $x>0$ while $\dot{T}>0$ for $x<0$.  This pair of
S-branes seems to be necessary to create a D0-brane on a
non-BPS D1-brane.

Now we can define our charge conservation rule.  If we just
consider the D0-brane and the $\overline{\rm D0}$-brane, charge
is not conserved.  To conserve charge we must include the S-brane
charge and so propose the following conservation law.  For any
closed curve, for example the dashed circle in the figure, count
the number of D-branes and S-branes which flow into the curve in
such a way that a D-brane (anti-S-brane) contributes a charge
$+1$ while an anti-D-brane (S-brane) counts as a $-1$. Naturally,
a single stationary D0-brane conserves charge as does a single
flat S0-brane (which is consistent with the charge conservation
of the known flat S-branes of the rolling tachyon.) In the above
figure the net change inflow is zero, $+2-2=0$.

The verification of this conservation law is straightforward.
Draw an arbitrary simple closed curve over the spacetime plot of
any tachyon configuration and parametrize the curve by $l$,
so the values of the tachyon are $T=T(l)$, $0 \leq l\leq
2\pi$.  The zeros of the tachyon configuration are located at
$l= l_i$ where $i=1,2,\cdots,2n$.  Now the important
point is that we take the tachyon field to be a single valued
function over the worldvolume $T(l=0)=T(l = 2\pi)$, so integrating the
derivative $\partial T/\partial l$ over the
curve we get
\begin{eqnarray}
\sum_i \mbox{sgn}\left[\frac{\p T}{\p l}
\biggm|_{l=l_i}\right]=0. \label{topology}
\end{eqnarray}
The locations $l_i$ with $\mbox{sgn}\left[\frac{\p T}{\p
l}\bigm|_{l=l_i}\right]=+1$ are physically interpreted as
intersections of the circle with either a D0-brane or
$\overline{\rm S0}$-brane,
depending on how fast the tachyon field zeros are moving. This
proves our conservation law and clearly shows that S-branes play
an essential role in charge conservation.\footnote{More precisely,
the ``location'' $l_i$ does not specify the location of the branes
but gives the maximum of the RR charge density. The RR charge
density is given by $\sim e^{-T^2} dT$, and the integration over
$T \in [-\infty, \infty]$ gives a unit RR charge. In the
following the location should be understood in this sense of the
maximum charge density. }

\begin{figure}[tp]
\begin{minipage}{50mm}
\begin{center}
\includegraphics[width=5cm]{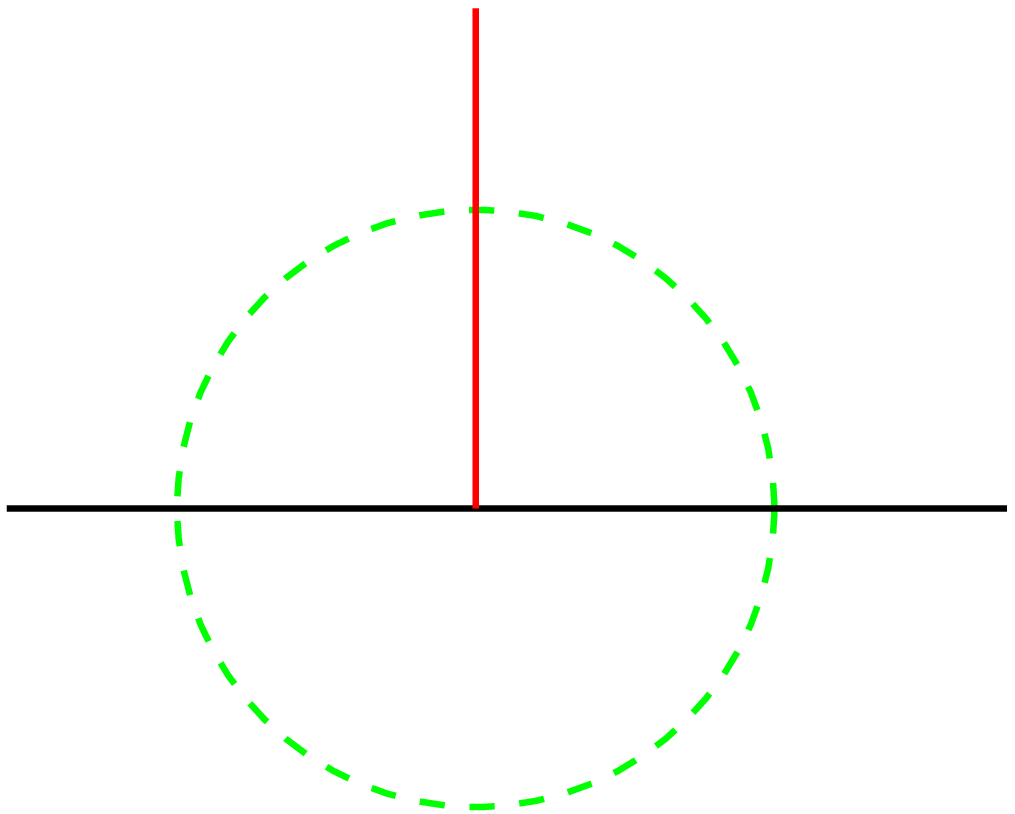}
\put(-110,50){S0?} \put(-70,100){$\overline{\rm D0}$} \put(-40,50){S0?}
\caption{Creating a $\overline{\rm D0}$-brane does not conserve charge.}
\label{chargeviolation}
\end{center}
\end{minipage}
\hspace*{5mm}
\begin{minipage}{50mm}
\begin{center}
\includegraphics[width=5cm]{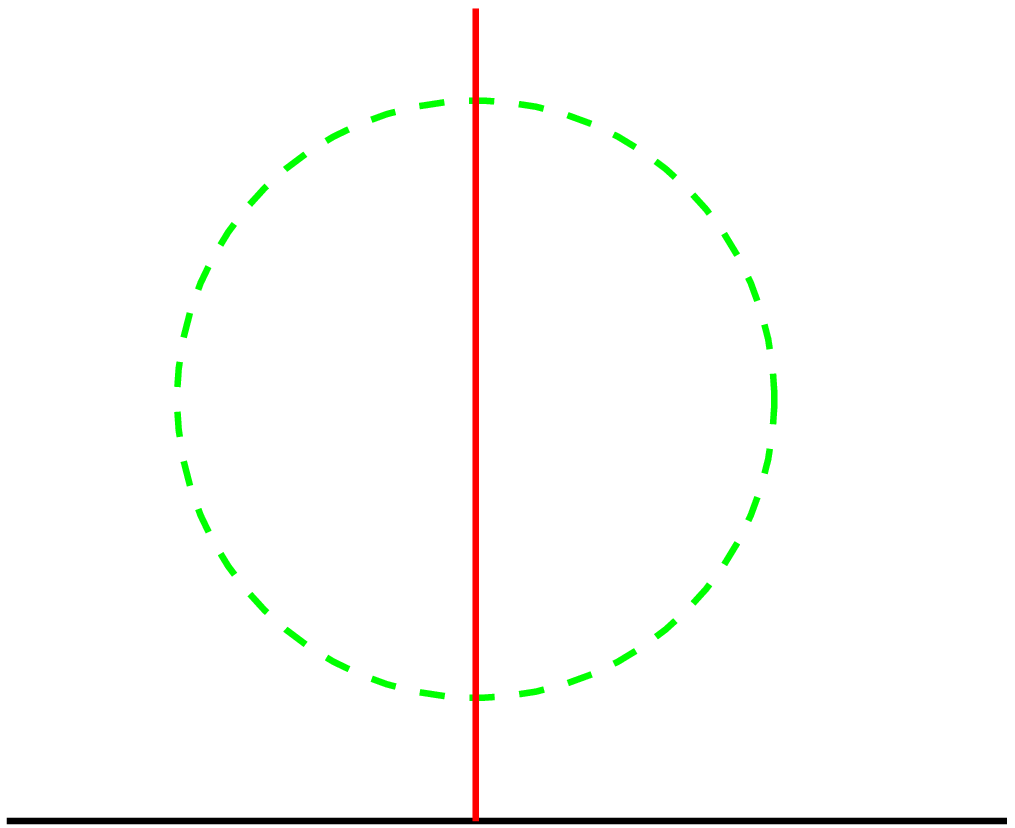}
\caption{Putting the S-brane at past infinity will ensure charge
conservation.} \label{chargeatinfinity}
\end{center}
\end{minipage}
\hspace*{5mm}
\begin{minipage}{50mm}
\begin{center}
\includegraphics[width=5cm]{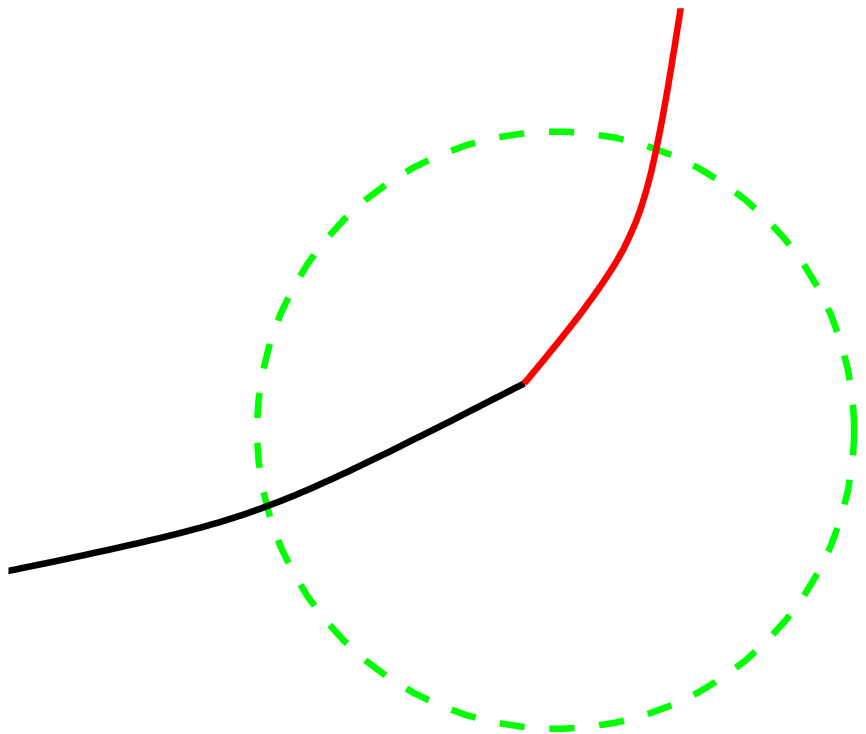}
\put(-140,40){S0} \put(-80,110){$\overline{\rm D0}$}
\caption{An S0-brane is
changing into a $\overline{\rm D0}$-brane. } \label{szerodzero}
\end{center}
\end{minipage}
\end{figure}

Consider next a similar case where the entire tachyon
configuration is situated at $T=0$. We are tempted to imagine the
formation of a net kink or anti-kink by tiny perturbations as
shown in Fig.~\ref{chargeviolation}, and this fact gives some
support to our previous statement that a flat S-brane is a good
candidate to describe the transition. Unfortunately this
observation is in direct contradiction to our charge conservation
law.  How do we resolve charge conservation with our above
observation? One way is to place the S-brane at past infinity by
reparametrizing time, see Fig.~\ref{chargeatinfinity}. The
S-brane can never be enclosed by any finite closed curve, so
charge is conserved. Putting the S-brane at past infinity was
also discussed in Refs.~\cite{Strotalk,LNT} as a ``half
S-brane'', where the tachyon was taken to be $T(t)=e^{\lambda
t}$.  This tachyon configuration is just like a flat S-branes in
our sense at early times and then dissipates into the vacuum at
late times. (To go from D0-brane to $\overline{\rm D0}$-brane we
would need something like $T(t,x)=x\sinh(\lambda t)$.) We may
also think of the situation illustrated in Fig.~\ref{szerodzero}
in which an S0-brane turns into a $\overline{\rm D0}$-brane so
charge is again conserved. Although charge conservation can not
solely determine the possible dynamics, it clearly does limit the
dynamical processes.

It should be remembered that we can produce chargeless
remnants.\footnote{Many field theories have solitons and so we
believe that Spacelike solitons (branes) should also exist in
these theories.  For example in scalar $\phi^4$ theory, it might
be possible for S-branes to describe the formation of the kink
solution.  In this case since the kink solution has $Z_2$
topological charge which should be conserved, the process
illustrated in Fig.~\ref{chargeviolation} still does not exist.}
The fundamental string formation process studied in section
\ref{sfs} provides an example. There the net (S-)brane RR charge
disappeared due to the shrinking worldvolume. Of course if we took
the branes to have zero charge then charge conservation would
play no role.  However as long as we treat topological defects
with topological charges, the same argument should apply.

Our discussion on charge conservation for codimension one kinks of
a real tachyon can be generalized to codimension two vortices of
complex tachyons, which exist on the worldvolume of a
D-$\bar{\textup{D}}$ pair.  Therefore in analogy to
Eq.~(\ref{topology}), the number of vortices and anti-vortices
intersecting a sphere should be equal.

Seeing how S-branes and D-branes interact, we remind of
string networks.  Also, one could attempt to interpret the
process in Fig.~\ref{charge1} as two copies of the process in
Fig.~\ref{szerodzero}.  Solutions of Fig.~\ref{szerodzero} are
not solutions of the S-brane action, but could be solutions of an
S-$\bar{\rm S}$ pair.


\section{Conclusions and Discussions}
\label{cad}

In this paper, we have explained how S-branes play a role in time
evolution in string theory, especially in the D-brane/F-string
formation during tachyon condensation. In general we have classified
S-brane solutions according to their remnants as in
Fig.~\ref{remnantchart}. Although there are some ``expected''
solutions which we have not yet obtained, the arrows in
Fig.~\ref{remnantchart} typically show how S-branes work in
regards to time evolution of string theory processes. Although
our S-brane is defined through the rolling tachyon on non-BPS
D-branes, we may expect that this scenario of D-brane/F-string
formation via S-branes is more general and may be applied to
other situations of brane creation in string theory and also
to brane cosmology \cite{branecosmo}.  Possibly we
may even apply these S-brane methods to understand defect
formation in non-stringy systems with topological defects, such
as the standard model, since it has recently been reported that
the generic features of D-branes can be reconstructed in the
context of usual field theories \cite{Dvali}.

To illustrate the roles of the S-branes, we presented several
classical solutions of S-brane actions including:
electric S3-brane spike solutions (Section \ref{sfs} and
Ref.~\cite{Sbraneaction}) which described fundamental string
formation, electric-magnetic S3-brane spike solutions in Section
\ref{dfs} which produced $(p,q)$ strings and D-strings, and ``boosted''
S-branes which are flat and timelike branes capturing the late time
configuration of the spike solutions.  By directly analyzing the
non-BPS tachyon system in Section \ref{doc}, the confinement of
electric flux was shown to minimize the energy of the
corresponding tachyon system, and this result agrees with our
interpretation of the electric spike solution.  S-duality on the
S3-brane was studied in Section \ref{sis}, which turned out to be
consistent with the rolling tachyon with electric and magnetic
fields obtained in Section \ref{tswhemf}.  By taking into account
M-theory effects, we found out how to produce D-strings from an
S3-brane.  The existence of these solutions therefore
demonstrates that S-duality could in fact be used in a new way to
constrain remnant formation.  Our resolution of the imaginary
field strength on the S3-brane worldvolume is potentially
relevant in other cases \cite{Hull}.
The boosted S-brane was introduced and we provided their
corresponding tachyon configurations in Section \ref{tcr}. We
also obtained the boosted S-brane boundary state which clarified
that the boosted S-brane is T-dual in the time direction to
$(p,q)$ strings.  In our analysis the fundamental string limit of
$(p,q)$ strings can be described by both D-branes and S-branes
so the critical electric field $E=1$ is likely a self-dual point
between these two descriptions.

We now turn to detailed comments on some results we obtained in
this paper. Although the late time configuration of the spike
solution in Section \ref{sfs} is given by the boosted S-brane in
Section \ref{bsas}, we haven't found explicit tachyon solutions
corresponding to the spike solutions of Section \ref{sfs} and
Section \ref{dfs}. The results of Ref.~\cite{hirano}, which
discussed tachyon spike configurations of D-branes (the
brane/F-string ending on branes), might be useful in the
construction of tachyon configurations for S-branes. It might be
possible to generalize the recent result in Ref.~\cite{Senrecent}
on the correspondence between the tachyon system and DBI on their
defects, to our S-brane situations. It is inevitable, however,
that the tachyon solutions will be approximate since the precise
Lagrangian in string theory is still missing.  Also, while work
has been done to check various static properties of tachyon
actions, their time dependent properties are not as well
understood.

We also point out various other solutions and generalizations.
Another type of solution to look for on the S-brane worldvolumes
we have discussed, is to have the electric field and magnetic
fields in different directions. One example is to have the
electric field along the $\chi$ direction and to have the
magnetic field along one of the angular directions, let us say
$\phi$.  A similar static case has been discussed in
Refs.~\cite{supertube, tube-elemag}.   Also, in the solution of
Section \ref{sfs}, it is possible to take a T-duality along the
$\chi$ direction. This simply turns $F_{t\chi}$ into the velocity
along that direction, so the criticality of the original electric
field will result in the S-brane worldvolume moving at the speed
of light. This is a null geodesic, and looks like an emission
process of a D-brane. Another interesting generalization is to
have multiple spikes. This is possible because the BIon spike
solutions in Refs.~\cite{CallanMaldacena, Gibbons} decouple from
each other and so do the multiple S-brane spikes. These solutions
are similar to the above emission processes. In this case we
observe many D-branes and strings coming in from past infinity and
scattering to various directions in the target space. However we
would like to state that such a configuration is odd since
although the Hamiltonian is simply the sum of spikes, and hence
gives seemingly independent worldvolumes, we see that the
worldvolumes also apparently intersect for some time.

The analysis in Section \ref{sis} also implies that there are also
throat solutions in S-brane systems as in the D-brane cases. In
the ordinary D-brane case the throat solutions are relevant for
the brane and anti-brane annihilation process \cite{CallanMaldacena,
HK-Decay, J7}. It would be very interesting if the role of these
S-brane throat solutions (the throat is along time direction
$X^0$ in the S-brane case) is clarified. In fact this question is
related to the possible non-Abelian structure of S-branes which
should be not just the result of non-Abelian structure of the
original non-BPS D-branes but is more intrinsic to time evolution
and tachyon condensation on a single non-BPS D-brane. Since the
throat can also carry electric charge, it is possible that these
throat solutions are involved with the mechanism of electric flux
confinement.

Finally, the various S-brane solutions we have found are
reminiscent of interactions between branes and strings, and the
interpretation that particular S-brane solutions can be thought
of as Feynman diagrams was pursued partly in Section \ref{sadi}
and Ref.~\cite{J7}. In Section \ref{sadi} the creation of
codimension one D-branes was qualitatively discussed from the
viewpoint of charge conservation. We believe that this creation
process can be described by some classical solution of the
(tachyonic) S-brane action which might be the action of an
S-$\bar{\rm S}$ pair. However here we outline another possible
way to describe this D-brane creation process. The tachyon
configuration of Fig.~\ref{charge1} has an interesting property.
The rolling tachyon energy at $x\neq 0$ is nonzero while at $x=0$
the energy is equal to the non-BPS brane since $T=\dot{T}=0$
there. The tension of the S0-brane depends on the rolling tachyon
energy ${\cal E}(x)$, since in the derivation of the S-brane
action, the tension of the S-brane is just the value of the
non-BPS brane Lagrangian integrated over $x^0$ with substitution
of the classical solution $T(x^0)$ which is dependent on ${\cal
E}$. Hence it may be possible to regard ${\cal E}(x)$ (or
equivalently, the tension of the S-brane) as another dynamical
variable that the S-brane system has. At values of $x$ with
${\cal E}(x) = {\cal T}_{\rm non-BPS D1}$, a $\overline{\rm
D0}$-brane is created as in Fig.~\ref{charge1}. If we may
introduce a term like $(\partial {\cal E})$ in the S-brane
Lagrangian, it may fix the spatial dependence of the S-brane
tension via equations of motion for ${\cal E}$ and so govern the
D-brane creation process. However since ${\cal E}(x)$ is not a
localized mode on the S-brane but defined through the integration
over all the $x^0$ region, it might be difficult to proceed along
this direction to generalize the S-brane Lagrangian.

If a configuration like Fig.~\ref{charge1} is explicitly
constructed, however, it should provide an interesting procedure
to compute Feynman diagrams for D-brane scattering. It is possible
that physical quantities associated with the scattering process
are directly related to S-brane actions and their solutions.
Understanding these S-brane systems might provide a theory of
interacting D-branes and strings in a general context and an
alternative to Matrix theory.

By analyzing boundary states with electric fields and an
inhomogeneous tachyon background, the authors of
Ref.~\cite{SJR-Sug-emergent} have also recently discussed
solutions which can dynamically produce fundamental strings. It
would be interesting to further explore the relationship between
their boundary state analysis and S-brane solutions.

We leave these issues to future investigation.


\section*{Acknowledgements}
We thank C.~-M.~Chen, J.~Evslin, M.~Garousi, M.~Kleban, J.~Kumar,
H.~-N.~Li, M.~Li, N.~Sakai, Y.~Sakamura, A.~Sen, S.~Sugimoto and
S.~Terashima for valuable discussions and comments.
K.\ H.\ is supported in part by the Grant-in-Aid for Scientific
Research (No.\ 12440060 and 13135205) from the Japan Ministry of
Education, Science and Culture.
P.\ M.\ H. and J.\ E.\ W.\ are supported in part by
the National Science Council,
the Center for Theoretical Physics
at National Taiwan University,
the National Center for Theoretical Sciences,
and the CosPA project of the Ministry of Education, Taiwan, R.O.C.


\appendix
\section{Tachyonic S-brane Action}
\label{appa}

In this appendix, we explicitly demonstrate how the tachyonic
S-branes considered in Section \ref{vortex} appear in the tachyon
condensation of D-brane anti-D-brane. In Fig.~\ref{descentt},
this is the arrow (2). Since the arrow (1) has already been
discussed in Ref.~\cite{Sbraneaction}, while the arrow (3) is just the
same as the usual D-brane descent relation, and the arrow (4) was
realized in Section \ref{vortex}, the derivation of (2) completes
the explanation of the generalized descent relations of
Fig.~\ref{descentt}.

To derive the effective action of the ``tachyonic S-brane'' by
using the fluctuation analysis of the time dependent kink as
performed in Ref.~\cite{Sbraneaction}, we return to the Lagrangian of
the D-\Dbar\ pair in Eq.~(\ref{BSFT}) and the solution
representing  the tachyonic S-brane in Eq.~(\ref{solts}). A
direct analysis of this fluctuation mode is difficult due to the
complexity of the Lagrangian.

The easiest way to proceed is to simplify the situation and
truncate the derivatives of the Lagrangian at fourth order
\begin{eqnarray}
S={2T_{\rm  D9}} \int d^{10} x \ e^{-|T|^2} \left(1+  |\p_\mu
T|^2 + p_1 \left(|\p_\mu T|^2\right)^2 +p_2 (\p_\mu T)^2(\p_\nu
\bar{T})^2 \right) ,
\label{dertru}
\end{eqnarray}
where $T\equiv T_1+iT_2$ and $p_1, p_2$ are numerical constants.
We must keep at least fourth order derivative terms since if we
only keep the usual canonical kinetic energy there are no tachyon
solution linear in time, and unless the solution is linear it is
again technically difficult to perform a fluctuation analysis.
The equation of motion for a homogeneous time dependent
tachyon is
\begin{eqnarray}
- T_1 (1+\dot{T}_1- 3 p \dot{T}_1^4) +\ddot{T}_1(1-6 p
\dot{T}_1^2)=0 ,
\end{eqnarray}
where $p\equiv p_1+p_2$, and we have set $T_2=0$. Therefore the
linear solution
\begin{eqnarray}
T_{\rm{cl}}=a x^0 , \label{linearsolu}
\end{eqnarray}
exits for $a=\sqrt{\frac{1+\sqrt{1+12p}}{6p}}$.

It is actually strange that we have completely linear solution in
spite of the presence of the tachyon potential. The higher-order
kinetic term makes this situation possible. The general solution
does not exhibit the rolling tachyon behavior at late time, since
this model is just a generalization of the Minahan-Zwiebach model
which does not possess the rolling tachyon behavior. The general
solution reaches the true vacuum $T=\infty$ in finite time. But
if we tune the initial condition then we have the completely
linear solution for the rolling tachyon. The strangeness of this
solution is also apparent in that its energy vanishes
\begin{eqnarray}
 {\cal E} =
\int e^{-|T|^2} (1+\dot{T}_1- 3 p \dot{T}_1^4)=0.
\end{eqnarray}
For the meantime we treat this special solution as just an
illustration of the new descent relations.

\vspace{5mm}

\noindent
\underline{Fluctuation spectrum}

\noindent
Let us consider the following fluctuation
\begin{eqnarray}
T=T_{\rm{cl}} (x^0) +t_1 (x^\mu) +i t_2(x^\mu),
\end{eqnarray}
where $\mu = 0,1,\cdots,9$. Substituting this into the action and
collecting terms quadratic in the fluctuation fields, we obtain
the fluctuation action
\begin{eqnarray}
S_{\rm{fluc}}=&&{}\hspace{-20pt}
{2T_{\rm  D9}}\int\! d^{10}x \; e^{-T_{\rm{cl}}^2}
\left[
(\frac{4}{3}-\frac{2}{3} a^2)\big((2a^2 x_0^2 -1 )t_1^2
-t_2^2\big)
-\frac{8-4a^2}{3}x_0 t_1\dot{t}_1
\right.
\\
&&\hspace{10mm}
\left.+\frac{a^2-2}{3a^2}(\p_\mu t_1)^2+\frac{4(1+a^2)}{3a^2}
\dot{t}_1^2 +(1-2a^2p_1+2a^2p_2)(\p_\mu t_2)^2+4p_2a^2\dot{t}_2^2
\right].
\nonumber
\end{eqnarray}
The two fluctuation modes are completely decoupled from each
other.  Integrating by parts, we find that
\begin{eqnarray}
&& S_1  ={2T_{\rm  D9}}\int\! d^{10}x\; e^{-T_{\rm{cl}}^2} \left[
\frac{a^2-2}{3a^2} (\p_\mu t_1)^2+\frac{4(1+a^2)}{3a^2}
 \dot{t}_1^2
\right], \\
&& S_2 ={2T_{\rm  D9}}\int\! d^{10}x\; e^{-T_{\rm{cl}}^2} \left[
\left(\!-\frac{4}{3}+\frac{2}{3}a^2\!\right)
t_2^2 + (1-2a^2p_1+2a^2p_2)(\p_\mu
t_2)^2+4p_2a^2\dot{t}_2^2 \right] . \quad \hspace{5mm}
\end{eqnarray}
To see the physical meaning of these fluctuations, we redefine
the fields as $\hat{t}_{1,2}= e^{-(ax^0)^2/2} t_{1,2}$ so the
newly defined fields $\hat{t}_{1,2}$ have canonical kinetic
terms. Then we can decompose the fields $\hat{t}_{1,2}(x^\mu)$
into the eigenfunctions of the harmonic potential along $x^0$, as
performed in Ref.~\cite{Sbraneaction}.   We may determine the ``mass''
spectra for these fluctuations as the eigenvalues of the
Laplacian, $\partial_i^2$, for the spatial directions.

The $t_1$ fluctuation contains a zero mode which is the Nambu-Goldstone
mode
associated with the symmetry breaking of the translation along
$x^0$ by the presence of the kink solution. The `mass' tower of
$t_1$ is obtained as
\begin{eqnarray}
m^2=\frac{8a^2(1+a^2)}{a^2-2} n, \quad n=0,1,\cdots .
\end{eqnarray}
The constant $a$ should be less than $\sqrt{2}$ to keep the
coefficient of the term $(\p_\mu  t_1)^2$ negative.

Next, we use the field redefinitions to rewrite the action $S_2$
as
\begin{eqnarray}
S_2={2T_{\rm  D9}}\!\int\! d^{10}x
\left[(1\!-\!2a^2p_1\!+\!2a^2p_2)(\p_i \hat{t}_2)^2
+\frac{2\!-\!a^2}{3a^2}
\hat{\dot{t}}_2^2+\frac{a^2(2\!-\!a^2)}{3a^2}x_0^2 \hat{t}_2^2
+(a^2\!-\!2)\hat{t}_2^2 \right].
\end{eqnarray}
{}From this expression it is easy to extract the mass spectrum
\begin{eqnarray}
m^2=\frac{2-a^2}{3a^2(1-2a^2p_1+2a^2p_2)} \left[
a\sqrt{\frac{2-a^2}{3}}(2n+1)+a^2-2 \right] .
\end{eqnarray}
Here again $1-2a^2p_1+2a^2p_2>0$ should be satisfied so that the
fluctuation Lagrangian is positive definite. The lowest mode
becomes tachyonic, and this tachyonic mode is associated with the
instability of the time-dependent kink solution.

\vspace{5mm}

\noindent
\underline{Effective action}

\noindent The lowest modes in the fluctuations $\hat{t}_{1,2}$
are Gaussian, and if one expresses these in term of the original
fluctuation then they are actually constant, independent of
$x^0$. Using this property, we can calculate the effective action
for the tachyonic S-brane. By substitution of the fluctuation
into the original D-\Dbar\ action, we have
\begin{eqnarray}
S\!\!&&={2T_{\rm  D9}}\int\! dx_0 \;e^{-(ax_0+t_2)^2} \int d^9x
\Big[
1-a^2 +(\p_i
t_1)^2 +(\p_i t_2)^2
\bigg.
\nonumber \\
&& \hspace{5mm}
+p_1 \Big(a^4-2a^2(\p_i t_1)^2-2a^2(\p_i t_2)^2 \Big)
+p_2 \Big(a^4-2a^2(\p_i t_1)^2+2a^2(\p_i t_2)^2 \Big)
\Big]
\nonumber
\\
&&={2T_{\rm  D9}}\frac{\sqrt{\pi}}{a} \int d^9x e^{-t_2^2} \Big[
\frac{2}{3}(2-a^2) -\frac{2-a^2}{a^2} (\p_i
t_1)^2+(1-2a^2p_1+2a^2p_2) (\p_i t_2)^2
\nonumber
\\
&& \hspace{35mm}
+\big((\p t)^4\rm{term}\big) \Big].
\label{tsac}
\end{eqnarray}
(In the last line we have performed the integration over $x^0$.)
This is the tachyonic S-brane effective action, which resembles a
Minahan-Zwiebach model \cite{Zwie}.
The difference between them are as
follows: (1) the sign of $(\p_i t_1)^2$ term  is negative,
indicating that this mode represents the translation along the
time direction. (2) the worldvolume metric defining this theory
is Euclidean. These two properties are shared with the S-brane
action obtained in our previous paper.

Although we have adopted a derivative truncation as the starting
point (\ref{dertru}) and also a special solution
(\ref{linearsolu}), we believe that this effective action
(\ref{tsac}) may capture essential features of the tachyonic
S-branes.

\section{Evaluation of Tachyon Energy of Boosted S-brane}
\label{appb}
\setcounter{footnote}{0}

Though the energy of the (deformed) S-brane configurations has
been studied in Section \ref{sfs}, Section \ref{dfs}, and
Ref.~\cite{Sbraneaction}, the overall normalization of the S-brane
action has not been specified there. This can be fixed in
principle in the derivation of the S-brane actions in Section
\ref{vortex} and
Ref.~\cite{Sbraneaction}. It is clear that the factor
$S_0$ in Ref.~\cite{Sbraneaction},
which is an ``S-brane tension,'' can
be computed by substituting the rolling tachyon solution into the
original tachyon action.   This tension $S_0$ is not therefore
fixed since it is dependent on the rolling tachyon energy
${\cal E}$.
This situation is different from the case of static tachyon
defects of D-branes where the tension is fixed completely.

Let us evaluate $S_0$ using the BSFT Lagrangian as a starting
point.\footnote{So far, among many tachyonic Lagrangians, only
the BSFT Lagrangians reproduce the D-brane tensions correctly and
consistently.} The BSFT action of a non-BPS D2-brane is\footnote{We have
rescaled the tachyon from that of Ref.~\cite{BSFTmatter} as
$T \rightarrow T/\sqrt{4\pi}$. }
\begin{eqnarray}
 S_{\rm nonBPS} = -{\cal T}_2 \int d^3x \; e^{-\pi T^2}
\sqrt{-\det(\eta + F)} {\cal F}\left(z\right),
\label{tacacnd2}
\end{eqnarray}
where the worldvolume coordinates are $x^0, x^1, \chi$ and
\begin{eqnarray}
 z \equiv G^{\mu\nu}\p_\mu T \p_\nu T.
\end{eqnarray}
We are working in the units $2\pi\alpha'=1$, and $G_{\mu\nu}$ is
the open string metric. The function ${\cal F}$ is defined by
BSFT and its explicit form is given in
Refs.~\cite{BSFT,BSFTsoliton,BSFTRR} for example. The properties of
this ${\cal F}$,
\begin{eqnarray}
 {\cal F}(z) \sim -\frac12 \frac1{z+1} \quad (z \sim -1)
\label{expansion}
\end{eqnarray}
will turn out to be important later.

For vanishing field strength the homogeneous rolling tachyon
solution $T=T_{\rm cl}(x^0)$ presented in Ref.~\cite{BSFTmatter} has
an asymptotic expansion for large $x^0$
\begin{eqnarray}
 T_{\rm cl}(x^0)= x^0 + \epsilon(x^0)
+ \mbox{higher},
\end{eqnarray}
where
\begin{eqnarray}
 \dot{\epsilon}(x^0) =
\sqrt{\frac{ {\cal T}_2 }{{4}{\cal E}}}
\exp\left[-\frac{\pi}{2}(x^0)^2\right].
\end{eqnarray}
Here ${\cal E}$, the energy density of the above homogeneous
rolling tachyon solution, is defined by the following Hamiltonian
density formula
\begin{eqnarray}
 H =  {\cal T}_2 \;e^{-\pi{T_{\rm cl}}^2}
\sqrt{-\det(\eta + F)} \left[
{\cal F}(z) - \dot{T} \frac{\delta z}{\delta \dot{T}}
\frac{\delta {\cal F}(z)}{\delta z}\right].
\label{densityformula}
\end{eqnarray}
Note that $T_{\rm cl}(x^0)$ is a function dependent
on the integration constant ${\cal E}$ implicitly. The S-brane
tension $S_0$ is just the value of the action (\ref{tacacnd2})
into which the solution $T_{\rm cl}$ is substituted (while the
integration over the spatial worldvolume is left unperformed, to
give the worldvolume of the S-brane). Although the complexity of
the function ${\cal F}(z)$ obstructs the analytic evaluation of
the action, we can read off the integrand in the asymptotic
region $x^0\sim\infty$. Noting that $z$ approaches $-1$ in this
limit
\begin{eqnarray}
 z \sim -1 -
\sqrt{\frac{ {\cal T}_2 }{\cal E}}
\exp\left[-\frac{\pi}{2}(x^0)^2\right],
\end{eqnarray}
we obtain
\begin{eqnarray}
 L_{\rm nonBPS}
\sim -{\cal T}_2\;
e^{-\pi (x^0)^2}\left(-\frac12\right)
\left[
-\sqrt{\frac{ {\cal T}_2 }{\cal E}}
\exp\left[-\frac{\pi}{2}(x^0)^2\right]
\right]^{-1}
=
-\frac{\sqrt{{\cal E}{{\cal T}_2 }}}{2}
e^{-\pi (x^0)^2/2}.
\end{eqnarray}
This means that the value of $S_0$, which is given by the
integral of $L_{\rm nonBPS}$ over $x^0$, is in fact finite and
may be approximated as
\begin{eqnarray}
 S_0 \sim
-\frac{\sqrt{{\cal E}{{\cal T}_2 }}}{2}
\int_{-\infty}^{\infty} \! dx^0\;e^{-\pi (x^0)^2/2}
=- \sqrt{\frac{{\cal E}{{\cal T}_2}}{2}}.
\label{stension}
\end{eqnarray}

Let us move on to the evaluation of the energy of the boosted
S-brane which is a timelike object. It is straightforward to show
that the rolling tachyon solution in the presence of a constant
magnetic field is also a solution of the non-BPS D2-brane system
(\ref{tacacnd2}),
\begin{eqnarray}
 T = T_{\rm cl}(x^0), \quad F_{1\chi}={\rm const.}
\end{eqnarray}
Basically we can turn on the constant field strength transverse
to the S-brane freely. Next, consider the boosted solution
\begin{eqnarray}
 T = T_{\rm cl}\left({x^0}'\right), \quad F_{1\chi}= \mbox {const.}
\label{boostsolution}
\end{eqnarray}
where
\begin{eqnarray}
 {x^0}' \equiv x^0 \cosh \gamma  + x^1\sqrt{G_{11}} \sinh\gamma.
\label{locS}
\end{eqnarray}
Here the open string metric is
\begin{eqnarray}
 G_{\mu\nu} = {\rm diag}(-1,1+F_{1\chi}^2, 1+F_{1\chi}^2).
\end{eqnarray}
One can show that (\ref{boostsolution}) is again a
solution\footnote{The nontrivial check is
on the equations of motion for the gauge fields. The tachyon
equation of motion is trivially satisfied since we made a boost
respecting the open string metric.} of the non-BPS D2-brane
system (\ref{tacacnd2}).  In the limit
\begin{eqnarray}
 F_{1\chi} \rightarrow \infty
\label{Flimit}
\end{eqnarray}
the S-brane becomes timelike and in this case the tachyon
configuration is approximately
\begin{eqnarray}
 T \sim \left( \sqrt{G_{11}} \sinh \gamma\right) x^1,
\end{eqnarray}
which resembles
the usual D-string kink solution.
This suggests that the energy is localized at $x^1=0$.

We keep this in mind and proceed to carefully evaluate the
Hamiltonian at $x^0=0$ for simplicity. The asymptotic expansion
of $T_{\rm cl}$ at $x^0=0$ is
\begin{eqnarray}
 T =  \sinh \gamma \sqrt{G_{11}} x^1 +
\sqrt{\frac{{\cal T}_2}{4{\cal E}}} \exp \left[
-\frac{\pi}{2}(\sinh\gamma \sqrt{G_{11}} x^1)^2 \right]+ {\rm
higher}
\end{eqnarray}
and this approximation is very good for nonzero $x^1$ and large
$F_{1\chi}$. For this solution the argument $z$ is
\begin{eqnarray}
 z = \left(
-\dot{T}^2 + G^{11} (\partial_1 T)^2
\right)
= ... = - \left( T_{\rm cl}'\right)^2
\end{eqnarray}
where ${'}$ denotes a derivative with respect to the argument of the
function $T_{\rm cl}$, i.e. in the above
\begin{eqnarray}
 T_{\rm cl}' \equiv \left[
\frac{\delta T_{\rm cl}(a)}{\delta a}
\right]_{a = \cosh \gamma x^0 + \sinh \gamma \sqrt{G_{11}} x^1}.
\end{eqnarray}
Since $T'_{\rm cl}$ approaches $1$, $z$ approaches $-1$ everywhere
except $x^1=0$ in the limit $F_{1\chi} \rightarrow \infty$. This means
that in the evaluation of the energy $\delta{\cal F}/\delta z$
(the second term in (\ref{densityformula})) is
much larger than ${\cal F}$ (first term in (\ref{densityformula})) due
to the expansion (\ref{expansion}),
so the Hamiltonian at
$x^0=0$ is given by
\begin{eqnarray}
 H \hspace{-15pt}&&={\cal T}_2 \exp\left[
-\pi (\sinh\gamma\sqrt{G_{11}} x^1)^2
\right]
\sqrt{-\det (\eta + F)}\;
2(\dot{T})^2 \frac{\delta {\cal F}(z)}{\delta z}
\nonumber\\
&& =
{\cal T}_2 \exp\left[
-\pi(\sinh\gamma\sqrt{G_{11}} x^1)^2
\right]
\sqrt{1\!  +\! F_{1\chi}^2}
\; 2(\cosh^2\gamma) ({T}_{\rm cl}')^2
\frac{\delta {\cal F}(z)}{\delta z}
\nonumber\\
&& =
{\cal T}_2 \exp\left[
-\pi (\sinh\gamma\sqrt{G_{11}} x^1)^2
\right]
\sqrt{1\!  +\! F_{1\chi}^2}
\; 2(\cosh^2\gamma)
\frac{1}{\frac{2{\cal T}_2}{\cal E} \exp\left[
-\pi (\sinh\gamma \sqrt{G_{11}}x^1)^2
\right]}\nonumber \\
&&
= {\cal E}\sqrt{1+F_{1\chi}^2} \cosh^2\gamma.
\label{bakene}
\end{eqnarray}
This is independent of $x^1$, and we have shown that the background
rolling tachyon energy is still present everywhere even in the limit
$F_{1\chi} \rightarrow \infty$.
(The above result is consistent with the original rolling tachyon with
$F_{1\chi}=0$ and $\gamma=0$, since this should give the energy
${\cal E}$.)

Let us consider higher order terms in the Hamiltonian to see the
localization of the energy which should correspond to the energy of the
boosted S-brane.
In the limit (\ref{Flimit}), it turns out
that the next-to-leading order term coming
from the expansion of the potential term $e^{-\pi T^2}$
can be ignored.
First, we expand the function $z$ for large $x^1$ at
$x^0=0$ as
\begin{eqnarray}
 z = -1 - \sqrt{\frac{{\cal T}_2}{\cal E}}
\exp \left[
-\frac{\pi}{2} (\sinh\gamma \sqrt{G_{11}} x^1)^2
\right]
-\frac{{\cal T}_2}{\cal E}
\exp \left[
-\pi (\sinh\gamma \sqrt{G_{11}} x^1)^2
\right]
+ \mbox {higher}.
\end{eqnarray}
Then the Hamiltonian is
evaluated to the next-to-leading order as
\begin{eqnarray}
 H &=& {\cal E}\sqrt{1+F_{1\chi}^2} \cosh^2\gamma
\left[1\;\;-\;\;\frac12 \sqrt{\frac{{\cal T}_2}{{\cal E}} }
\exp \left[
-\frac{\pi}{2} (\sinh\gamma \sqrt{G_{11}} x^1)^2
\right]
\right]
\nonumber
\\
&&
\hspace{10mm}
+ \frac12 \sqrt{{\cal ET}_2}
\sqrt{1+F_{1\chi}^2}
\exp \left[
-\frac{\pi}{2} (\sinh\gamma \sqrt{G_{11}} x^1)^2
\right]
+ \mbox{higher.}
\end{eqnarray}
Here
the second term in the first
line is from the higher order evaluation of $\delta {\cal F}/\delta z$
in (\ref{densityformula}), while
the second line comes from evaluation of ${\cal F}(z)$ term in the
Hamiltonian (\ref{densityformula}).
Interestingly, though these two exponential terms
become infinitely small in the limit $F_{1\chi}\rightarrow \infty$,
they are combined and approach a $\delta$ function whose coefficient is
finite. More
precisely, the above expression is arranged in this limit as
\begin{eqnarray}
 H
= {\cal E}\sqrt{1+F_{1\chi}^2} \cosh^2\gamma
-\sqrt{\frac{{\cal E}{\cal T}_2}{2}}|\sinh\gamma| \delta(x^1).
\label{resultenergy}
\end{eqnarray}
So, in addition to the homogeneous energy of the background rolling
tachyon, we have a localized energy with a finite coefficient!
This second term should be identified with the energy of the boosted
S1-brane.

We now show that the localized energy contribution we just
calculated agrees with the Hamiltonian of the S1-brane action.
The action of a static S1-brane located at $x^1=0$ is
\begin{eqnarray}
 S_{\rm S1} = S_0 \int\! dx^0 dx^1 d\chi \;
 \delta (x^1)\sqrt{E^2-1}.
\end{eqnarray}
Using this action, one finds that
the S1-brane Hamiltonian density is
\begin{eqnarray}
 H_{\rm S1} = S_0 \frac{1}{\sqrt{E^2-1}}\; \delta (x^1).
\label{SH}
\end{eqnarray}
Now this electric field $E$ is the induced electric field as seen
in Eq.~(\ref{inducedE}). After taking the limit
$F_{1\chi}\rightarrow \infty$, we have $ E = -\coth \gamma$.
Substituting this into the S-brane Hamiltonian (\ref{SH}), we
obtain
\begin{eqnarray}
 H_{\rm S1} = S_0|\sinh\gamma|\;\delta (x^1).
\end{eqnarray}
Remarkably this agrees with the finite energy contribution in
Eq.~(\ref{resultenergy}) using the S-brane tension of
Eq.~(\ref{stension})!

Lastly we provide a comment on this localized energy. In the final
expression (\ref{resultenergy}), the S-brane contribution was
found to be negative. This suggests that the S-brane has a
negative energy, which agrees with the result of the boundary
state analysis in which the time-time component of the boosted
S-brane boundary state is given by a negative value as opposed to
the usual boundary states for $(p,1)$ strings.  In this appendix
we have shown why this does not result in any of the usual
problems.  While the contribution of the S-brane is negative,
there is an additional leading order energy contribution in
Eq.~(\ref{resultenergy})
which is due to the energy of the background
rolling tachyon,
and so the total energy is still positive.

The picture is reminiscent of anti-particles in the ``Dirac sea''.
The boosted S-brane is like something existing in a cloud of
fundamental strings. Since our non-BPS D-brane formulation did
not take care of the radiation of the fundamental strings, it
keeps the energy and effect of all these strings which are
supposed to radiate away. (One of the effects of this cloud of
fundamental strings might possibly be to make the S-brane
energy negative.)
Actually, the string cloud will dissipate, and the S-brane with
strings attached to it will become a D-brane with strings
attached to it.
(Here we have to distinguish the strings on
the non-BPS D-brane which will decay away,
from strings stuck to the S-brane.)
As a final remark,
the energy of the background rolling tachyon
in (\ref{bakene}) diverges in the limit
$F_{1\chi}\rightarrow \infty$. The validity of
some of the calculations are not so rigorous
due to this singular limit. Although the boosted S-brane is expected to
capture the late time behaviour of the spike solution in
Section \ref{sfs}, apparently this divergence
comes from the fact that we have
not taken into account the curved worldvolume of the S-brane in the
spike solution where $F_{1\chi}$ is divergent only at $r=0$. In this
sense the correspondence between the spike solution and the boosted
S-brane is not exact.

\newcommand{\J}[4]{{\sl #1} {\bf #2} (#3) #4}
\newcommand{\andJ}[3]{{\bf #1} (#2) #3}
\newcommand{\AP}{Ann.~Phys.~(N.Y.)}
\newcommand{\MPL}{Mod.~Phys.~Lett.}
\newcommand{\NP}{Nucl.~Phys.}
\newcommand{\PL}{Phys.~Lett.}
\newcommand{\PR}{ Phys.~Rev.}
\newcommand{\PRL}{Phys.~Rev.~Lett.}
\newcommand{\PTP}{Prog.~Theor.~Phys.}
\newcommand{\hep}[1]{{\tt hep-th/{#1}}}

\end{document}